\documentclass[12pt,preprint,round,colon,authoryear,longnamesfirst]{aastex}

\begin{document}

\shortauthors{Metchev, Hillenbrand, \& Meyer}
\shorttitle{Ten-micron Observations of Young Stars}

\title{Ten-micron Observations of Nearby Young Stars}
\author{Stanimir A.\ Metchev, Lynne A.\ Hillenbrand}
\affil{California Institute of Technology}
\affil{Division of Physics, Mathematics \& Astronomy, MC 105-24, Pasadena, CA 91125}
\email{metchev, lah@astro.caltech.edu} 
\and
\author{Michael R.\ Meyer}
\affil{The University of Arizona}
\affil{Steward Observatory, 933 North Cherry Avenue, Tucson, AZ 85721-0065}

\received{}
\revised{}
\accepted{}
\journalid{}{}
\articleid{}{}

\begin{abstract}
We present new 10$\micron$ photometry of 21 nearby young stars obtained at the
Palomar 5-meter and at the Keck~I 10-meter telescopes as part of a
program to search for dust in the habitable zone of young stars.
Thirteen of the stars are in the F--K spectral type range (``solar
analogs''), 4 have B or A spectral types, and 4 have spectral type M.
We confirm existing {\sl IRAS} 12$\micron$ and ground-based 10$\micron$
photometry for 10 of the stars, 
and present new insight into this spectral regime for the rest.  
Excess emission at 10$\micron$ is not found in any of the young solar analogs,
except for a possible 2.4-sigma detection in the G5V star 
HD~88638.  The G2V star HD~107146, which does not
display a 10$\micron$ excess, is identified as a new Vega-like candidate,
based on our 10$\micron$ photospheric detection, combined with previously 
unidentified 60$\micron$ and 100$\micron$ {\sl IRAS} excesses.
Among the early-type stars, a
10$\micron$ excess is detected only in HD~109573A (HR~4796A),
confirming prior observations; among the M dwarfs, excesses are confirmed in
AA~Tau, CD~--40$\degr$8434, and Hen~3--600A.  A previously suggested 
$N$ band excess in the M3 dwarf CD~--33$\degr$7795 is shown 
to be consistent with photospheric emission.

We calculate infrared to stellar bolometric luminosity ratios for 
all stars exhibiting
mid-infrared excesses, and infer the total mass of orbiting dust in the
cases of optically thin disks.  For a derived median photometric
precision of $\pm$0.11~mag, we place an upper limit of $M_{\rm
dust}\approx2\times10^{-5}M_\earth$ on the dust mass (assuming a dust 
temperature of 300~K) around solar analogs not seen in excess at
10$\micron$. Our
calculations for the nearby K1V star HD~17925 show that it may have the least 
massive debris disk known outside our solar system ($M_{\rm
dust}\gtrsim 7\times10^{-6}M_\earth$).  

Our limited data confirm the expected tendency of decreasing fractional 
dust excess $f_d=L_{\rm IR}/L_*$ with increasing stellar age.  However, we 
argue that estimates of $f_d$ suffer from a degeneracy between the 
temperature and the amount of circumstellar dust $M_{\rm dust}$, and 
propose a relation of $M_{\rm dust}$ as a function of age, instead.
\end{abstract}

\keywords{circumstellar matter --- dust, extinction --- planetary systems: 
protoplanetary disks --- infrared: stars}

\section{INTRODUCTION}

Since the discovery of far-infrared (far-IR) excess emission by 
the {\sl Infrared
Astronomy Satellite (IRAS)} toward the main sequence star Vega \citep{aum84},
nearly 400 other main-sequence stars have been 
detected with such excesses \citep[][and references therein]{man98,syl00,
hab01,spa01,lau02a,son02}, interpreted as due to orbiting dust.
Too old to possess remnant primordial dust (which would be cleared
by Poynting-Robertson drag on the time-scale of several Myr in the
absence of gas), these stars owe their far-IR excess to emission by
``debris disks,'' formed
by the continual collisional fragmentation of larger bodies \citep[][and
references therein]{bac93}.  Stellar systems exhibiting this phenomenon
have been accordingly named ``Vega-like.''
Subsequent higher spatial resolution imaging at millimeter wavelengths 
of the nearest subsample of Vega-like stars has revealed intricate 
disk-like structures, with gaps and concentrations 
indicative of ongoing disk clearing by planets
\citep[e.g., ][]{hol98, gre98, jay98, koe98, wil02}.  

While such observations have
provided a wealth of information about the outer reaches ($>10$~AU) of these
systems, the ``habitable zone'' (defined as the region corresponding to
the temperature of liquid water) around the
majority of them remains at separations un-resolvable by
direct imaging, and can be probed only by
interferometry or by mid-infrared (mid-IR) spectroscopy.  
To first order, however,
information about the presence of warm ($\sim$300~K) material around a
star can be extracted from its spectral energy distribution (SED), and the
presence or absence of a mid-IR excess above the expected photospheric
level.

To date, searches for mid-IR excesses from both the ground and space 
have been dominated by upper limits.  All Vega-like stars
have been identified from {\sl IRAS} and {\sl Infrared Space
Observatory (ISO)} observations,
and are therefore limited by their sensitivities
($\sim$0.5~Jy at 12--60$\micron$ for the {\sl IRAS} Point Source Catalog).
In a sample of nearly 150 stars in the 1--630~Myr
age range, \citet{spa01} find that the vast majority of targets older
than $\sim$20~Myr have either no measurable 12 and 25 $\micron$ excesses 
or no detections from {\sl IRAS}.  Nevertheless, given a gradually 
decreasing dust fraction for young stars \citep{spa01, hab99},
more sensitive mid-IR observations are expected to uncover
less massive debris disks around older stars.

Young stars form a particularly suitable set
of targets to search for mid-IR emission from circumstellar disks.
They are likely to be at a transition
stage, where the innermost regions of the dust disk have been
cleared, whereas, due to an increase of the viscous and dynamical timescale
with radius \citep{hol00,nak83}, cooler
material may still be present farther away.  No excess is expected at 
near-infrared (near-IR) wavelengths (1--5$\micron$; $\lesssim 0.1$~AU)
from these stars, yet longer-wavelength excess due to more 
distant (0.1--10~AU), cooler material, could be detectable.

Except for stars within a few tens of parsecs, stellar photospheres
are not detected at 12$\micron$ with the small apertures of {\sl IRAS} 
and {\sl ISO}.  However, the recent availability of mid-IR
instrumentation on 5--10-meter class ground-based telescopes has made them
competitive with {\sl IRAS/ISO} within the atmospheric windows at 10$\micron$ 
and 20$\micron$, with limiting flux densities $\lesssim$20~mJy detectable 
at 10$\micron$.  In particular, for sufficiently nearby and
sufficiently young stars, sensitivity to stellar photospheres is
achieved.  This implies sensitivity to a small excess above the
photosphere, and hence to debris disks $\sim 10^{-5} M_\earth$, limited
by photometric precision.  In
addition, the higher angular resolution attainable with large
ground-based telescopes allows for unambiguous determination of the
origin of the infrared (IR) flux within the {\sl IRAS/ISO} beam.  We can thus
discriminate between Vega-like excesses associated with the star, and
apparent excess flux associated, instead, with a background object in the {\sl
IRAS} or {\sl ISO} beam \citep{aum91,lis02}, structure in the IR cirrus, 
or due to detection threshold bias \citep{son02}.

In this paper, we present $10\micron$ photometry of 21 nearby stars
spanning the $\sim$1--600~Myr age range, and thus expected to cover
the stages in the transition between a primordial and a debris disk.
Age estimates are obtained from the literature, and are based on 
high-resolution spectroscopic measurements of
lithium abundances, and/or from inferred association with known 
kinematic groups 
of stars.  Our sample was selected to include nearby ($\lesssim150$~pc) 
stars, covering a wide range of spectral types (B9--M4), for which the
photosphere should be detectable at 10$\micron$ from the ground.

Twelve of our targets have {\sl IRAS} and/or {\sl ISO} detections, 
8 of which have already been inferred to harbor debris disks.  
For these objects, our 10$\micron$ data test the existence of 
possible excess at shorter wavelengths.

\section{GROUND-BASED TEN-MICRON PHOTOMETRY}

Observations were conducted under marginally photometric conditions at 
$10.7\micron$ 
with the Long Wavelength Spectrograph \citep[LWS;][]{jon93} on the Keck~I
10-meter telescope during two runs (2000 February~20--21 and December~9), and
under photometric conditions at $10.3\micron$ with SpectroCam-10 
\citep[SC--10;][]{hay93} on the Hale 
5-meter telescope, on 2002 January~1 and 3.  The $10.7\micron$ LWS
filter has a full-width half-maximum (FWHM) of
$\approx1.6\micron$\footnote{http://www2.keck.hawaii.edu/inst/lws/filters.html}, and the $10.3\micron$ SC--10 filter has a bandwidth of
$\sim1\micron$ \citep{hay96}, and are thus both longward of the telluric
ozone feature at 9.6$\micron$, and of the peak of the potential photospheric 
silicate emission at 9.7$\micron$.

Table~\ref{tab_dates} lists the instrument and
epoch of observation for each source.

\subsection{LWS Data}

LWS is a Boeing-made 128$^2$~pix Si:As moderate-flux
array, mounted at the Forward Cassegrain module on the Keck~I
telescope, providing diffraction-limited imaging ($10\arcsec$ field 
of view, 0.08$\arcsec$/pixel) and spectroscopic
($R$=100 and 1400) capabilities in the 3.5--25$\micron$ wavelength
range.  For the observations presented here we used only the imaging 
mode of LWS at $10.7\micron$.  The seeing was poor and variable: 
0.3$\arcsec$--0.6$\arcsec$ at 10$\micron$ (the attainable diffraction
limit is 0.22$\arcsec$).  
Data were taken in the standard manner
employed for mid-IR observations, with both chopping and nodding
to off-source positions.  The resulting images were 
processed and co-added
using the {\sc LWSCOADD} routine written in {\sc IDL} by Gregory D.\ 
Wirth.\footnote{Available at
http://www.astro.caltech.edu/mirror/keck/inst/lws/lwscoadd\_pro.html}
No flat-fielding was performed, as this was found to increase the
scatter in the photometry.  Instead, we attempted to place the
stars always in the same place on the array.

Magnitudes were obtained based on measurements of 10 IRTF $N$-band standard 
stars \citep{tok84,rie85}.
Photometry was performed with the {\sc IRAF/PHOT} task in apertures of 
radius 0.96$\arcsec$ (12~pix), with sky background measured in 
2.0$\arcsec$--3.2$\arcsec$-radius (25--40~pix) annuli.  A curve-of-growth
correction ($-0.12\pm0.01$~mag, based on standard star measurements) to a
1.92$\arcsec$ (24~pix) aperture was 
then applied to each star to bring the magnitudes to within $\approx$2\% of
the infinite aperture value.
An atmospheric extinction correction of $-0.32\pm0.07$~mag/airmass 
was derived based on standard star observations, where we have
allowed for $\approx$20\% shifts in the night-to-night atmospheric zero-points 
to align the extinction curves from each night.  The root-mean-square (r.m.s.)
scatter in the photometry of the standards is 0.06~mag for 2000 February 
20 and 21, and 0.13~mag for 2000 December 9.

\subsection{SC--10 Data}

SpectroCam-10 is a 128$^2$~pix Si:As BIBIB (Back Illuminated Blocked
Impurity Band) detector manufactured by Rockwell.  For our program we
used it in imaging mode ($16\arcsec$ field of view) with a $10.3\micron$
filter, although the instrument also allows $R$=100 and 2000 spectroscopy 
between 5 and 20$\micron$.  The seeing was poor, 0.7$\arcsec$ at 10$\micron$,
but relatively stable.  Both chopping and nodding were used in the
data acquisition, with the chop and the nod throws being of the same 
magnitude ($\approx$20$\arcsec$).  The chip was
binned into a 64$\times$64~pix array with 0.256$\arcsec$ pixels,
providing nearly Nyquist sampling of the diffraction limit (0.48$\arcsec$)
of the telescope, and then stored into a 3-dimensional image, each slice
of which contained a co-added chopped frame.  Dark frames were taken for all
source exposure times, and subtracted from all frames.  Flat fields 
were constructed by median-combining the off-source exposures for each target.
This approach of obtaining a flat field
was preferred to making a master flat from all exposures during the night, 
because of an unexplained $\sim$4$\arcsec$
shift in the filter vignetting pattern across the field of view over the
course of the observations.  The r.m.s.\ scatter in the flat fields was
$\lesssim$1\%.

Final images from SC--10 were obtained by differencing each nodded pair,
and then averaging the differences.  Since source re-acquisition during
nodding was not perfect, the PSF of the final image was broader 
(FWHM~$\approx3$~pix~$=0.77\arcsec$) than the diffraction limit.  Attempts
with centroiding the target in each differenced pair before co-adding were 
not successful, because only the standard stars had enough signal-to-noise
($S/N$) to obtain good centroids.  Object magnitudes were measured on the 
reduced images in 1.5$\arcsec$-radius (6~pix) apertures, using the {\sc
IRAF/PHOT} task.  Local sky was estimated as the mode of the pixels in
a 3.8$\arcsec$--$5.1\arcsec$ (15--20~pix) annulus around each object.
An aperture correction ($-0.147\pm0.014$~mag, determined from 
measurements of standard
stars) to a $3.1\arcsec$ (12~pix) aperture was then applied to bring 
the estimate of 
the enclosed flux to within $\approx$2\% of its infinite aperture value.
In the case of the double source RE~J0137+18A/B ($1.7\arcsec$ separation)
magnitudes were obtained through PSF-fitting with {\sc DAOPHOT/ALLSTAR} 
\citep{ste87}, using
a $3.1\arcsec$-radius PSF fit to the standard star $\beta$~And,
observed immediately beforehand.  An airmass correction of 
$-$0.13$\pm$0.03~mag/airmass, as determined from observations of the
standard stars $\alpha$~Boo and $\beta$~Gem on 2002 January 3, was applied 
to all measured magnitudes.  The r.m.s.\ scatter in the photometry of
the standards is 0.015~mag for 2002 January 1, and 0.007~mag for January 3.

\section{SEDS AND PHOTOSPHERIC MODELS}

\subsection{Optical and Infrared SEDs}

SEDs for the observed stars are presented in Figure~1, with
our 10$\micron$ photometry plotted as solid circles.  $UBVRILMNQ$ measurements
were collected from the literature, where available.  $JHK_s$ photometry
was extracted from the 2MASS Point Source Catalog for all but the
brightest sources (HD~216803, HD~102647, HD~17925, and HD~109573A), 
the data for which were taken from \citet{aum91} and \citet{jur93}.  
For the $UBVRI$ and $JHKLMNQ$ sets of filters, 
conversion from magnitudes to flux
densities $F_{\lambda}$ was performed using the zeroth-magnitude flux 
densities listed in \citet{all73} and \citet{coh92}, respectively.  For
the objects with 2MASS photometry, the transformation from $JHK_s$
magnitudes to flux densities was based on the calibration by \citet{coh03}, 
outlined in the 2MASS Explanatory Supplement \citep[][ch.VI.4a]{cut03}.
For the $N$ filter we have adopted the same conversion factor
both for the 10.3$\micron$ (SC--10) and for
the 10.7$\micron$ (LWS) filters.  Given the difference in central
wavelengths of the two $N$-band filters, the derived zero-magnitude flux 
densities are expected to be within $\approx$4\% of each another.  

{\sl IRAS} fluxes and upper limits were 
obtained from the {\sl IRAS} Faint Source Catalog (FSC), or from the Point
Source Catalog (PSC) when the FSC did not contain the object
(HD~147809, HD~109573A, and the 100$\micron$ data point for 
CD~--40$\degr$8434).  FSC photometry was given preference over the PSC 
because of its higher signal-to-noise, and superior sensitivity 
to faint objects.  Reduced {\sl ISO} data in the 12--60$\micron$ range
were available for 5 of our
targets.  For HD~17925 and HD~102647 we adopted the color-corrected
25$\micron$ {\sl ISO} flux densities presented in \citet{lau02a};
the 11.5$\micron$ and 20$\micron$ data for HD~216803 are from
\citet{faj99}, and 60$\micron$ measurements for HD~145519 and HD~147809 
were taken from M. Meyer et al., in prep.
The {\sl ISO} Data Centre lists that additional data
were obtained for 5 of our targets (HD~147809, HD~145519, HD~216803, 
CD~--40$\degr$8434, and AP~93), none of which were user-reduced, however.
For the sources with no available reduced {\sl IRAS} or {\sl ISO} data we 
have adopted nominal {\sl IRAS} all-sky survey detection limits away from
confused regions of the sky: 0.5~Jy at 12, 25 and
60$\micron$, and 1.5~Jy at 100$\micron$.

The {\sl IRAS} and {\sl ISO} flux densities were color-corrected
assuming blackbody emission, following the procedures outlined in the
{\sl IRAS} Explanatory Supplement \citep[][ch.VI.C.3]{bei88} and the {\sl
ISO} Handbook \citep[][vol.IV, Appendix C]{lau02b}.  The adopted color
correction factors were chosen depending on the effective temperature $T_{\rm
eff}$ of each star (based on spectral types from Simbad, unless
otherwise noted; Table~\ref{tab_phot}).
For {\sl IRAS} they were 1.41--1.45 at 12$\micron$,
1.39--1.41 at 25$\micron$, 1.30--1.32 at 60$\micron$, and 1.09 at
100$\micron$, where the range of values corresponds to the range of 
$T_{\rm eff}$ of the stars in our sample.

Near- and mid-IR photometry for the observed sources from 2MASS
($JHK_s$), {\sl IRAS} (12$\micron$), and from our 10$\micron$
observations is summarized in Table~\ref{tab_phot}.
Twelve-micron color-corrected magnitudes ([12]) were calculated assuming an
{\sl IRAS} color-corrected flux of 28.3~Jy
for a zeroth-magnitude star \citep[][ch.VI.C.2a]{bei88}.

\subsection{Photospheric and Dust Models}

The stellar SEDs in Figure~1 were fit with
photospheric models from \citet{kur79,kur92} and with NextGen models from
\citet{hau99}.  
In fitting model photospheres, we have assumed $\log g=4.5$ and solar 
metallicity in all models, as is approximately true for main-sequence 
dwarfs in the solar neighborhood \citep{dri00}.  
The temperatures of the models were chosen to correspond
to the spectral types of the stars \citep[Table~\ref{tab_phot};][]{dej87}.  
This produced satisfactory (by eye) fits in most cases, except when 
non-simultaneous photometry of variable stars in different filters 
had to be fitted (e.g., AA~Tau, AP~93).  For RE~J0137+18A/B we
found that the required photospheric temperature ($T_{\rm
eff}\approx4700$~K for spectral type K3V) was too high to fit the optical
and near-IR SED.  We obtain better fits using a NextGen 4200~K
photosphere for both stars, which can be explained if both components
of the binary had sub-giant gravities \citep[$\log g \approx4.0$;][]{dej87}.
Given their young age ($\sim1$~Myr), AA~Tau, and DI~Tau are also
expected to have sub-giant gravities.  However, the effect on the model
photospheres is small compared to that due to the extinction measured toward 
both stars ($A_V=1.0$ and 0.76, respectively; Figure~1b).

The agreement between the two
sets of models is satisfactory over the 5000--10000~K range of
temperatures, though at lower temperatures the NextGen models describe
the stellar photosphere more closely \citep{hau99}.  For stars with $T_{\rm
eff}\geq5500$~K, we have chosen to plot
\citeauthor{kur92} models, as their lower resolution gives a better visual
idea of the integrated flux density of the stellar photosphere at short
wavelengths.  At wavelengths $\lambda>20\micron$, for which
\citeauthor{kur92} does not predict fluxes, we have linearly extrapolated 
the model photospheres from the last two (10$\micron$ and 20$\micron$) 
data points.  NextGen models (ranging from 10~nm to 10~mm) have been 
used for all cooler ($T_{\rm eff}\leq5000$~K) stars. 

The model for every star was normalized to each of the measured $J$, $H$ and 
$K$ fluxes (although in Figure~1 we have shown only the
$J$-band normalizations).  The mean predicted photosphere at 10.3$\micron$ 
(for the SC-10 targets) or at 10.7$\micron$ (for the LWS targets)
from the 3 normalizations was then taken as representative of the 
photospheric level, and the standard deviation of 
the mean~-- as the error ($\sigma_N{\rm pred}$) in the predicted photosphere.
The near-IR was chosen for normalization because: (1) it
is close in wavelength to the target band at 10$\micron$; (2) uniform 
photometry is 
available for the majority of our targets (from 2MASS); and (3)
most of our targets were not expected a priori to possess near-IR excesses.

We have assumed no circumstellar
extinction except in the cases for which an estimate was found in the
literature: HD~145519 \citep[$A_V=1.0$,][]{kre83}, HD~109573A
\citep[$A_V=0.03$,][]{jur91}, HD~147809
\citep[$A_V=1.6$,][]{egg98}, AA~Tau \citep[$A_V=1.0$,][]{bec90}, 
AP~93 \citep[$A_V=0.31$,][]{ode94}, DI~Tau \citep[$A_V=0.76$,][]{sta01}, 
and CD~--40$\degr$8434 \citep[$A_V=0.74$,][]{gre02}.  
An extinction law was adopted from \citet{mat90} over the 
entire SED range (0.25$\micron$--10mm).

Table~\ref{tab_phot} lists the derived 10$\micron$ excesses $\Delta N$ for
all targets.  The quoted error in $\Delta N$ is comprised in each case of the
photometric error in the measurement of $N$, and the error $\sigma_N{\rm
pred}$ in the level of the predicted photosphere.
Excluding the stars already known to possess such
(HD~109573A, CD~--40$\degr$8434, AA~Tau, and Hen~3--600A), and AP~93
(for which we only measure an upper limit to the 10$\micron$
intensity), the mean $N$-band excess is $0.01\pm0.03$~mag.  This is
indicative of the accuracy with which the photospheric models are able
to predict the broadband 10$\micron$ flux density, and demonstrates
that the models do not systematically over- or under-estimate the stellar
photosphere at $N$ band.

Excess IR emission, whenever apparent in the SED of a star, was
fit by a single blackbody curve, as would be expected from a narrow 
ring of dust at a
fixed distance from the star.   The temperature of the blackbody curve 
($T_{bb}$)
is identified for each case in Figure~1.  The fits were
performed by eye, and experimentation with a range of values for
$T_{\rm bb}$ showed that the determined
temperatures were generally accurate to $\pm$10\%.  For 
CD~--40$\degr$8434, AA~Tau, and Hen~3--600A, single temperature blackbody 
emission could not account
for the excess at wavelengths $>$60$\micron$, and for these stars we 
have allowed for additional, cooler blackbodies ($T_{\rm bb1}, T_{\rm bb2}$,
etc).  Although a disk model would
be more appropriate for fitting the observed excesses, our simplistic
treatment allows us
to estimate color correction factors for the {\sl IRAS} and {\sl
ISO} data (which were then re-applied instead of the ones based on the stellar
$T_{\rm eff}$), and to obtain approximate measurements of the dust luminosity
fraction in each case.

HD~17925 and HD~145519, with significant excesses only at 60$\micron$,
are the only stars for which we have not applied any color
corrections at wavelengths $\geq60\micron$.
The large uncertainty in the blackbody temperatures (30--250~K for
HD~17925, 30--140~K for HD~145519) 
that fit the far-IR excesses and upper
limits for these stars correspond to a range of values for
the color correction (0.91--1.19 for {\sl IRAS} at 60$\micron$ for
HD~17925, 0.93--1.08 for HD~145519).  
Thus the plotted (Figure~1) {\sl IRAS} and {\sl ISO} flux densities and upper 
limits for these
stars at wavelengths $\geq60\micron$ correspond to flat spectrum sources
($F_\nu \propto \nu^{-1}$), as is the default calibration for {\sl
IRAS} and {\sl ISO}.

\section{RESULTS}

\subsection{Stars with Confirmed Excesses}

Ten of our sample stars have been previously reported to possess mid- or
far-IR dust excesses: HD~145519, HD~109573A (HR~4796A), HD~147809,
HD~102647 ($\beta$~Leo), HD~17925, AA~Tau, DI~Tau, CD~--40$\degr$8434, 
Hen~3--600A, and CD~--33$\degr$7795.  Three among these
(HD~145519, HD102647, and HD~17925) have known
60$\micron$ dust emission from {\sl ISO} and/or {\sl IRAS}
\citep{bac97,lau02a,hab01}, but do not exhibit significant excesses 
at 10$\micron$, in agreement with previous measurements.  A fourth, DI~Tau,
previously suspected of harboring warm circumstellar dust on the basis
of 12$\micron$ {\sl IRAS} data \citep{skr90}, has subsequently been
shown not to possess
an excess at 10$\micron$ \citep{mey97}, and that all of the excess {\sl
IRAS} flux can be attributed to the neighboring (15.1$\arcsec$) DH~Tau.
Our photometry of DI~Tau is consistent with that of \citet{mey97} and
with the lack of circumstellar dust.  

Of the remaining 6 stars that {\sl have} been reported previously
to possess excesses at 10$\micron$ or 12$\micron$, we confirm
such in 4 of them: HD~109573A \citep{jur98}, AA~Tau \citep{str89}, 
CD~--40$\degr$8434 \citep{gre92}, and Hen~3--600A \citep{jay99a}. 
Our $N$-band data for these 4 objects
agree very well with the previously published 10$\micron$ photometry, as
well as with the color-corrected {\sl IRAS} 12$\micron$ flux densities,
and we will not discuss their photometry further.  

\subsection{Stars with Unconfirmed Excesses}

There are 2 stars~--- CD~--33$\degr$7795, and HD~147809~--- for which our 
non-detection
of excess at 10$\micron$ is in disagreement with previous data and
interpretations.

\subsubsection{CD~--33$\degr$7795}

\citet{jay99b} report a possible 2.6-sigma detection of a 10$\micron$ 
excess in the M3V star CD~--33$\degr$7795.  Their claim is based on a 
measured flux 
density of 96$\pm$9~mJy, and an estimated photospheric level of 70$\pm$5~mJy 
at 10.8$\micron$, assuming $K-N=0.0$.  Our 10.7$\micron$ 
flux density (93$\pm$8~mJy) is consistent with theirs, and with the level of
photospheric emission (100$\pm5$~mJy) found from the NextGen model.
The predicted $K-N$ color by the NextGen model is 0.27, which
agrees with the measured $K_s-N=0.24\pm0.04$ for this star.  For comparison,
\citet{ken95} list $K-[12]=0.49$~mag for M3V main-sequence
stars, which is expected to be comparable to their $K-N$ color.
There is thus no indication of
excess emission at 10$\micron$ in this object.  
\citeauthor{jay99b} acknowledge that if $K-N\sim0.3$ their measurement 
would be consistent with being photospheric.  

\subsubsection{HD 147809}

Our 10.7$\micron$ measurement of 46$\pm$6~mJy for this star, while
consistent
with being photospheric, is a factor of 5 fainter than the {\sl IRAS}
flux density of 243$\pm24$~mJy at 12$\micron$.
HD~147809 is detected at 12$\micron$ and 60$\micron$ ($S/N\approx5$ in
both bands), but is listed only in the PSC, despite being fainter than 
the nominal PSC sensitivity at these wavelengths ($\sim$0.5~Jy), and
brighter than the sensitivity limit of the FSC ($\sim$0.2~Jy).
However, the {\sl IRAS} Faint Source Catalog
Rejects (FSCR) lists a 12$\micron$ source (IRAS~Z16219-2514) within 
5$\arcsec$ of
HD~147809 (associated with IRAS~16219-2514): a separation
equal to a fifth of the angular resolution of {\sl IRAS} at 12$\micron$.
The reason for rejection (and inclusion in the FSCR) is
that IRAS~Z16219-2514 is
detected only in 1 band (12$\micron$) in the FSC, with signal-to-noise of 
3--6 \citep[][actual $S/N=5.9$]{mos92}.  The FSCR also lists one
nearby 100$\micron$-only cirrus extraction (``cirrus'' flag 1), 
which may have possibly affected the PSC 60$\micron$ measurement.   The
possibility that the 60$\micron$ and the 12$\micron$ PSC 
detections are associated with IR cirrus is explored below.

HD~147809 is a member of the
Upper Scorpius OB complex in the $\rho$~Oph region, which is associated 
with highly-variable background emission in the mid- and far-IR
\citep{ryt87}, as seen in the {\sl IRAS} maps of this area.  
\citet{bac97} observe the star with {\sl ISO} at 60$\micron$ and
90$\micron$ with a detection only at 60$\micron$:
$0.91\pm0.07$~Jy ($S/N$=13)~--- 4-sigma fainter than the
$1.64\pm0.18$~Jy flux listed in the {\sl IRAS} PSC at this wavelength.
This is possible if the peak of the 60$\micron$ source was in the 
central (45$\times$45$\arcsec$) pixel of the 3$\times$3 PHT--C100 
{\sl ISO} array, and if the emission extended over the 8 border pixels, which
were used to estimate local background.
{\sl IRAS} photometry, on the other hand, does not involve local 
background
subtraction \citep[each survey scan is initiated and terminated with a 
flash of the internal reference source to monitor responsivity of the
system;][ch.VI.B.1.]{bei88}, and thus measures the total flux in the
{\sl IRAS} beam \citep[60$\arcsec$ at 60$\micron$;][ch.II.C.3.]{bei88}.
The lower {\sl ISO} flux density is therefore indicative of the extended 
nature (over several C100 pixels: $\gtrsim$50$\arcsec$) of the 
60$\micron$ emission.  At the distance of the Upper Sco OB association 
\citep[$\sim$160~pc;][]{deg89,jon70}, the physical size of the
structure would be
$\sim$10000~AU.  While it is possible that HD~147809 possesses
a dust disk of such size, the corresponding dust temperature
(10--30~K) is inconsistent with the 90$\micron$ {\sl ISO} upper
limit (read off from the default {\sl ISO} post-stamp measurement
representation\footnote{Available at the ESA ISO website, 
http://www.iso.vilspa.esa.es/ida/index.html})

If the 12$\micron$ excess in HD~147809 were associated with
orbiting circumstellar material around the star, the ratio of
the PSC 12$\micron$ and 25$\micron$ (upper limit) flux densities
$S_\nu(12\micron)/S_\nu(25\micron)\geq0.95\pm0.09$ would require a blackbody
temperature $T_{\rm bb}\geq154$~K (3-sigma lower limit), whereas the measured 
$S_\nu(10.7\micron) / S_\nu(12\micron) = 0.19\pm0.03$ requires 
$T_{\rm bb}=65\pm5$~K.  Therefore, the 12$\micron$ emission arises
either outside our 1$\arcsec$-radius LWS aperture,
or is extended over at least several arc seconds, s.t.
its intensity in the LWS aperture is comparable to that in the 
background annulus (2.0--3.2$\arcsec$).
An $S_\nu(12\micron)/S_\nu(25\micron)$ ratio greater than unity 
may be indicative of
non-thermal emission from aromatic infrared lines in the 12$\micron$ 
band, as previously suggested in association with inter-stellar cirrus in
$\rho$Oph \citep{mey93,ryt87}.  

Summarizing the evidence presented above, we conclude that the
12$\micron$ and 60$\micron$ excesses in the SED of HD~147809 are not
associated with the star, but are likely due to inter-stellar cirrus.

\subsection{New Excess Detections}

We find previously unknown IR excesses in 3 of our target stars.  
HD~88638 is the only one for which we present new evidence for
possible excess emission at 10$\micron$.  HD~107146 is an overlooked
Vega-like star from {\sl IRAS}, and HD~70516 has a 29$\arcsec$ distant
companion probably responsible for the 12$\micron$ {\sl IRAS} excess.

\subsubsection{HD~88638}

The flux measured from this star at 10.3$\micron$ is 2.4-sigma 
($\Delta N=0.19\pm0.08$~mag) above the 5500~K, solar metallicity and gravity
($\log g=4.5$) Kurucz photosphere.  
Given the \citet{hau99} claim that the NextGen models better approximate the
stellar photosphere at effective temperatures $\lesssim$5500~K, and the
spectroscopically determined $T_{\rm eff}=5360$~K for this star \citep{str00},
we also fitted
a 5400~K solar gravity and metallicity NextGen model to the SED.
The observed excess is 2.3-sigma over the NextGen model:
$\Delta N=0.18\pm0.08$~mag; i.e.\ the excess is nearly model-independent.

There is no measurement of the gravity or metallicity of HD~88638 in the
literature.  Given its {\sl Hipparcos} distance ($38\pm4$~pc) and magnitude 
($V=8.02$), and a bolometric correction of --0.165 \citep{flo96}, the star 
should have a luminosity of $L_*=0.83\pm0.15 L_\sun$, and a radius of 
$R_*=1.07\pm0.10R_\sun$.  Assuming $\sim 1M_\sun$ for a G5 star of
this luminosity, its gravity is $\log g=4.4$, i.e., nearly solar.
Alternatively, if sub-solar metallicity were causing the higher
10$\micron$ flux, it would have to be 
$\leq-2.0$~dex for the Kurucz model photosphere to fall within 1$\sigma$ 
of our $N$ band measurement.  
Such low metallicity would imply that HD~88638 is a halo 
(population~II) star, whereas its space 
velocity \citep[$U,V,W)=-(62.6,8.5,-1.3$;][]{str00} places its orbit in the 
plane of the galactic disk (population~I).  Gravity and metallicity
effects are thus unlikely to be the cause for the observed excess.

Interstellar extinction of order $A_V\approx0.1$ could contribute to the 
abnormal mid-IR emission.  However, HD~88638 is well out of the 
galactic plane ($b=+53\degr46\arcmin$), and at 38~pc from the Sun
extinction along the line of sight is expected to be negligible.  
Moreover, reddened photospheric
models normalized to the near-IR tend to underestimate the optical flux
by $\sim$0.2~mag.  Hence, extinction is not the probable cause.

The blackbody temperature range that accommodates the excess, the {\sl
IRAS} upper limits, and the near-IR data is 130--1500~K. This
does not exclude a substellar companion of spectral type L6 or
later ($T_{\rm eff}\leq1500$~K).  For an L6 dwarf $M_K\gtrsim11.4$, and assuming
$K-N \approx K-L' \gtrsim1.3$ \citep{leg02}, at the distance of the primary
(distance modulus 1.4~mag) the potential companion would have $N\gtrsim11.5$.
The 10$\micron$ excess that this would produce in HD~88638 ($N=6.04$,
Table~\ref{tab_phot}) is $\Delta N\lesssim0.01$: far too faint to account for 
what is observed.

Therefore, if real, the 2.4-sigma $N$ band excess is most likely due to 
orbiting debris. However, from a statistical point of
view, a 2.4-sigma positive deviation should
occur in 0.82\% of all observations, if the errors are Gaussian.
Given 21 observed sources, a random 2.4-sigma excess will occur 
in 16\% of such sets.  There is a hence non-negligible probability that 
the observed 10.3$\micron$ excess is random.

Due to the
marginality of our detection, and to the large range of possible blackbody
temperatures, we have not re-applied the corresponding color-correction
factors to the {\sl IRAS} upper limits.
These would range from 0.90--1.36 at 25$\micron$ and from
1.08--1.29 at 60$\micron$ (the ones applied based on a 5500~K stellar SED are
1.40 and 1.32, respectively), and will have the effect of allowing a
$\sim$10~K cooler blackbody fit, i.e., 120~K.

\subsubsection{HD~107146}

Although we do not detect above-photospheric emission at 10$\micron$
from this source, we choose to discuss it here because it has not been
previously identified as a Vega-like star, despite 60$\micron$ and
100$\micron$ excesses measured by {\sl IRAS}.
Because its 12$\micron$ {\sl IRAS} flux is consistent with
being photospheric, and because of a non-detection at 25$\micron$, it
would have been omitted in most previous searches concentrating on the
12$\micron$ and 25$\micron$ {\sl IRAS} bands.  However, this 
object has also been missed
in the \citet{ste91} survey for IR excess emission from {\sl IRAS} 
12$\micron$ point sources at high galactic 
latitude ($|b|>25\degr$).  This survey selected stars with 12$\micron$ {\sl
IRAS} detections from the PSC that were also detected in at least one 
more {\sl IRAS}
band, and that had positional association with an SAO star.  
It was apparently tailored to discover excess IR emission in 
sources like HD~107146 (SAO~100038, $b$=77.04), detected in the
12$\micron$, and in the 60$\micron$ bands in the PSC. However,
the low signal-to-noise ratios of these detections, $S/N$=3.9 and 6.9,
respectively, provide a clue as to why the star may have been overlooked.  
No {\sl ISO} data exist for this source.

The detection of an excess in this star is not false, as has 
been suggested for other reported candidate Vega-like stars with low-$S/N$ 
excess in a single {\sl IRAS} band.  The misidentification in those cases is 
explained as a likely result of detection threshold bias \citep{son02}. 
In the case of HD~107146 however, the FSC (which contains higher 
signal-to-noise measurements for objects fainter than the PSC 
completeness limits) lists detections of the star in
3, rather than 2 bands.  The higher significance of the
detection at 60$\micron$ ($S/N$=20), and a 100$\micron$ ($S/N$=6) detection
make it
highly improbable that the excess is due to an upwards noise fluctuation.
Moreover, our ground-based measurement is consistent with the listed 
{\sl IRAS} 12$\micron$ flux density, and with being photospheric,
thus excluding the possibility that the {\sl IRAS} flux comes from an
unassociated background object in the {\sl IRAS} aperture.

The IR excess in this star can be fit by a 60$\pm$10~K blackbody.  See
\citet{wil03} for a further discussion of 
this source, including recent sub-millimeter observations.

\subsubsection{HD 70516}

The 12$\micron$ {\sl IRAS} detection of this star is listed with a
10-sigma significance in the FSC, and is 4.1-sigma above the NextGen
photosphere ($\Delta[12]=0.45\pm0.11$), whereas our 10.3$\micron$ 
measurement is photospheric ($\Delta N=0.07\pm0.09$).  
HD~70516 (HIP~41184; G0) is listed as a proper motion binary in the 
{\sl Hipparcos}
catalog, with a $\Delta V=1.5$ K0 companion (HIP~41181) 29$\arcsec$ away.  The
separation between the two components matches the resolution of {\sl
IRAS} at 12$\micron$ \citep[$\sim$0.5$\arcmin$;][ch.I.A.1.]{bei88}, and it is
therefore likely that light from both components contribute to the
measured 12$\micron$ flux density.  Given $V$ magnitudes of 7.7 and 9.2
for the two stars, and expected $V-[12]$ colors of 1.5 and 2.0,
respectively \citep{wat87}, the combined 12$\micron$ magnitude is 5.85,
corresponding to $\Delta[12]=0.35$.  Although a significant fraction
of the companion's light will fall outside the {\sl IRAS} beam at
12$\micron$, its vicinity and brightness are marginally
consistent with the observed
12$\micron$ excess, making a debris disk interpretation less likely.

The binary is not detected at longer wavelengths with {\sl IRAS}, because 
its expected combined flux density is below the quoted sensitivity limits.

\subsection{Null Excess Detections}

Six of the remaining 8 stars do not have 12$\micron$ measurements from
{\sl IRAS}: RE~J0137+18A/B, HD~60737, HD~70573, CD~--38$\degr$6968,
and AP~93.  The first 5 all have 10$\micron$ fluxes consistent with 
being photospheric.
AP~93 (observed with the Palomar 5-meter telescope) was not detected 
at 10$\micron$.
The 3-sigma upper limit of 18.8~mJy for this source is 11.0~mJy above the
model atmosphere, and we can thus exclude any excess emission at this
level.  

HD~77407 and HD~216803 have 12$\micron$ measurements from {\sl IRAS}.
Our 10$\micron$ photometry for them
is consistent with the {\sl IRAS} data, and
the models show that the 10--12$\micron$ emission is at 
the expected photospheric level.

\section{DUST PROPERTIES OF STARS WITH EXCESSES}

\subsection{Fractional Dust Luminosity $f_d$}

Table~\ref{tab_excesses} lists the observed dust luminosity fraction 
$f_d=L_{\rm IR}/L_*$ for all stars with inferred disks
(marked with ``Yes'' in the last column).  These
were obtained based on the fitted blackbody curves to the IR excesses.
HD~88638 and HD~17925 allow a range of temperatures in the blackbody fits, and
for them we have listed a range of values for $f_d$, with hotter blackbodies
generally producing higher fractional luminosities.  The measured values 
for $f_d$
correspond well to the ones found in the literature ($F_{d ({\rm Lit})}$; see
the last column of Table~\ref{tab_excesses} for references).  We are therefore
confident that our blackbody fits to the excess IR emission 
have not introduced significant systematic errors in the estimate of $f_d$,
with respect to estimates based on more elaborate dust and/or debris disk 
models \citep[e.g., as in][]{chi01,li03}.

For the sources without excess emission at 10$\micron$ we estimate 
$f_d \lesssim 10^{-3}$, approximately corresponding to the 
3-sigma upper limit on the dust excess that would be produced by a
hypothesized 300~K debris ring around a G5 star, assuming our median 
uncertainty of 0.11~mag in the measured excess $\Delta N$.  This is
appropriate, since most of the stars that we do not detect in excess 
are G and K dwarfs.  DI~Tau is the only M dwarf without a detected 
excess, and given its later spectral type, the upper limit on $f_d$ 
is somewhat higher: $f_d\leq3\times 10^{-3}$, whereas for HD~147809,
the only early-type star (A1V0 not seen in excess, $f_d\leq5\times 10^{-4}$.
For HD~216803, for which the SED is further constrained by the 
12$\micron$ and
25$\micron$ {\sl IRAS} photospheric detections, this upper limit is also
$f_d\lesssim5\times10^{-4}$, whereas for AP~93, for which we only have
an $N$-band upper limit, $f_d\lesssim7\times10^{-3}$.

\subsection{Dust Disk Parameters \label{sec_diskparams}}

In estimating the mass of the observed circumstellar dust disks, we 
have followed the formalism and the steps outlined in
\citet{che01}:

\begin{eqnarray}
a \geq a_{\rm min} = \frac{3L_* Q_{\rm pr}}{16\pi G M_* c \rho_s} \\
n(a)da = n_0 a^{-p} da \\
M_{\rm dust} \geq \frac{16\pi}{3} f_d \rho_s D^2 \langle a
	\rangle \\
t_{\rm PR} = \frac{4\pi \langle a \rangle \rho_s}{3} \frac{c^2
	D^2}{L_*} \\
M_{\rm PB} \geq \frac{M_{\rm dust}}{t_{\rm PR}} t_{\rm age} .
\end{eqnarray}

Equation (1) gives the minimum particle size $a_{\rm min}$ that is not
blown out by radiation pressure \citep{art88}; $L_*$ and $M_*$ are the 
stellar luminosity and mass, $Q_{\rm pr}$ is the radiation pressure coupling
coefficient, and $\rho_s=2.5$~g~cm$^{-3}$ is the mean 
density of a dust grain (as assumed in \citealt{che01}).
Equation~(2) represents the assumption that the particle size
distribution $n(a)da$ results from equilibrium between
production and destruction of objects through collisions
\citep{gre89}, where we have assumed $p=3.5$ \citep{bin00}.
If we weight by the number of particles, the equation can be
integrated to obtain an average particle size $\langle a \rangle = 5/3
a_{\rm min}$.  The inferred
dust mass $M_{\rm dust}$ around the star is found from Equation~(3), where we
have assumed that the dust is distributed in a thin shell at a
distance $D$ from the star \citep{jur95}.  If the grains are larger 
than $\langle a
\rangle$, or if colder dust is present ($\lesssim 30$~K; undetectable
by {\sl IRAS/ISO}), 
our estimate of $M_{\rm dust}$ is a lower bound.  The
Poynting-Robertson lifetime $t_{\rm PR}$ of circumstellar grains is given by
Equation~(4) \citep{bur79}.  Since this timescale is generally shorter 
than the ages ($t_{\rm age}$) of the stars in our sample 
(Tables~\ref{tab_phot} and \ref{tab_excesses}), collisions between larger
bodies are required to replenish the grains destroyed by PR drag.  The
mass $M_{\rm PB}$ of the parent bodies is given by Equation~(5).

Stellar parameters (radius $R_*$, mass $M_*$, and bolometric
luminosity $L_*$) for stars of spectral type K1 and earlier were
estimated
using the \citet{all99} database of fundamental parameters of nearby
stars,
given $T_{\rm eff}$ (based on spectral type) and assuming
$\log g=4.5$.  For the cooler ($T_{\rm eff}<5000$~K) stars these
values were taken from the
literature, or, in the case of Hen 3--600A, $L_*$ and $R_*$
were estimated using the star's
distance, its visual magnitude (assuming $A_V=0.0$), and
bolometric correction
from \citet{flo96}.  Table~\ref{tab_starpars} lists the parameters
adopted for each star.

The radiation pressure coupling coefficient $Q_{\rm
pr}$, which enters in the grain size calculation (Equation~1), was
assumed
$\approx$1, since it deviates from unity only for grains smaller than
$\lambda_p/2\pi$ \citep{bur79}, where $\lambda_p$ is the wavelength of
peak emission from the host star.  For the coolest K dwarf in our sample
(RE~J0137+18A/B, assumed $T_{\rm eff}=4200$~K)
$\lambda_p/2\pi=0.1\micron$, which
is comparable to the blow-out size for this star.  Our assumption that
$Q_{\rm pr}\approx1$ is therefore valid for K3 and earlier-type
stars.

In deriving the distance $D$ of the dust shell from the star we have
expanded the discussion in \citet{che01} by
adding a differentiation between grains with blackbody 
(BB; perfect emitters and absorbers), greybody (GB; imperfect emitters, 
perfect absorbers), and inter-stellar medium (ISM; imperfect emitters 
and absorbers) properties.  The three cases have been applied as
detailed in \citet{bac93}, and we have inverted their formulae (3),
(5), (6) to solve for $D_{\rm BB}, D_{\rm GB}$, and $D_{\rm ISM}$ 
(in units of AU), respectively:

\begin{eqnarray}
D_{\rm BB} = \left(\frac{278}{T_{\rm dust}}\right)^2 
	\left(\frac{L_*}{L_\sun}\right)^{1/2} \\
D_{\rm GB} = \left(\frac{468}{T_{\rm dust}}\right)^{5/2} 
	\left(\frac{L_*}{L_{\sun}}\right)^{1/2} 
	\frac{\lambda_0}{\micron}^{-1/2} \\
D_{\rm ISM} = \left(\frac{636}{T_{\rm dust}}\right)^{11/4}
	\left(\frac{L_*}{L_\sun}\right)^{1/2}
	\left(\frac{T_{\rm eff}}{T_\sun}\right)^{3/4}
\end{eqnarray}

In the relations above $T_{\rm dust}$ is the temperature of the dust grains
(corresponding to the temperatures $T_{\rm bb}$ of the excesses in
Figure~1), and $\lambda_0\sim
\langle a \rangle$ is the wavelength shortward of which the
radiative efficiency of a grain of size $\langle a \rangle$ is 
roughly constant: $\epsilon \sim (1-{\rm albedo})\sim 1$ \citep{bac97}.
Note however, that in fitting the excess in each star, we have only
used the blackbody law, without modifications at longer wavelengths
where particles of size $a<\lambda$ emit inefficiently.  The effect of
this is evident in the SEDs of HD~109573A and AA~Tau, where at
millimeter wavelengths the fitted blackbody overestimates the measured
flux.

The inferred dust properties of the detected circumstellar disks are
presented in Table~\ref{tab_disks}.

It is important to stress here that, given insufficient constraints from
the SED, there exists a degeneracy between the temperature ($T_{\rm
dust}$) and the amount of
material ($M_{\rm dust}$) around a star required to produce a certain
fractional excess $f_d$.  This is best seen in the case of HD~17925 in 
Figure~2.
The amount of 250~K debris ($M_{\rm dust}=7\times10^{-6}M_\earth$,
assuming GB particles) 
required to produce the upper limit on the fractional dust
excess in this star ($f_d=2\times10^{-4}$) is {\it
less} than what would be needed ($M_{\rm dust}=4\times10^{-5}M_\earth$)
to produce a smaller excess
($f_d=6\times10^{-5}$) at a cooler temperature (70~K).  
Yet, the two blackbody temperatures fit the
excess in HD~17925 comparatively well ($\chi^2 = 2.3$
and 2.5 per degree of freedom, respectively).  The best fit
($\chi^2_{\rm p.d.f.}=1.0$) is obtained for a 140~K dust producing a
fractional excess $f_d=7\times10^{-5}$, implying $M_{\rm
dust}=2\times10^{-6}M_\earth$.

The same trend is observed
in the fit to the SED of HD~88638, as discussed in 
Section~\ref{sec_88638} below.  We will revisit this
observation in Section~\ref{sec_fdmd} when considering the dependence 
of $f_d$ and $M_{\rm dust}$ on stellar age.

\section{DISCUSSION}

\subsection{Minimum Dust Masses of Stars with Excesses \label{sec_dustmass}}

\subsubsection{Known Optically Thick Disks: CD~--40$\degr$8434, 
AA~Tau, and Hen~3--600A}

The three coolest stars with disks, CD~--40$\degr$8434, AA~Tau, and 
Hen~3--600A
are young ($\lesssim10$~Myr), have high disk luminosity fractions 
($f_d\sim0.3$), and likely possess {\sl optically thick} disks.  All of 
them are, in fact, known T~Tauri stars.  Our estimated 
disk radii $D$ therefore are not physically meaningful.
Detailed modeling of T~Tauri disks is beyond the
scope of this paper, and we have not therefore attempted to estimate grain
sizes and disk masses for these three stars.  For in-depth discussion of
each case, see \citet[][CD~--40$\degr$8434]{gre02},
\citet[][AA~Tau; see also \citealt{bou99}]{chi01}, and 
\citet[][Hen~3--600A]{muz00}.  An estimate of the
total disk mass (dust+gas) exists only for AA~Tau \citep{bec90,chi01}, and we
have listed that in Table~\ref{tab_disks}.

For the remainder of the stars we have assumed that the disks are optically
thin, as expected given inferred dust luminosity fractions 
$f_d\lesssim0.01$, higher
host star temperatures ($T_{\rm eff}\geq5000$~K), and older ages
($\gtrsim10$~Myr).

\subsubsection{Known Optically Thin Disks: HD~145519, HD~109573A, 
HD~102647, and HD~17925}

Four stars have already been inferred to harbor debris disks, 
three of which (HD~109573A, HD~102647, and HD~17925) have been modeled 
in the previous literature.  For these we find that our estimates of the disk
parameters are in agreement with the existing ones.

\paragraph{HD~145519.}

HD~145519 (along with HD~147809) has been included in the analysis of
\citet{bac97} of {\sl ISO} excesses in open cluster A stars.  The star
possesses only a 60$\micron$ (3.0-sigma) excess and, 
unlike around HD~147809, there are no {\sl IRAS} reject sources or cirrus
extractions within a 5$\arcmin$-radius.  The {\sl ISO} excess is
therefore probably real.  The star is not discussed in
detail in \citet{bac97}, and no other measurements of the excess are
found in the existing literature.

\paragraph{HD~109573A (HR 4796A).}

Based solely on the aperture size that we have used for $N$ band photometry
(1$\arcsec$-radius for LWS), we can conclude that the
detected 10$\micron$ excess in HD~109573A 
arises within 70~AU of the host star.  This is consistent
with the extent of the disk predicted by the blackbody and greybody particle
assumptions, and excludes the possibility for the 
existence of smaller dust particles with properties similar to those of 
ISM grains.
Our estimate of the size of the debris disk around HD~109573A agrees with
previous mid-IR observations
\citep[][disk radii 40~-- 70~AU]{jay98,koe98,tel00}, 
and with the NICMOS scattered light images of \citet[][disk intensity 
peaking at 70~AU]{sch99}.  The dust mass derived here ($M_{\rm
dust}\gtrsim 0.83 M_\earth=5.0\times10^{27}$~g for GB particles) is
consistent with the one derived in \citet[][$3\times10^{26}$~g
$< M_{\rm dust} < 8\times10^{27}$~g]{jur95}, and with the 
estimates of \citet[][$M_{\rm dust} \approx 4.0-7.5\times10^{27}$~g]{li03} 
found from modeling of the near-IR scattered light and of the 
mid-IR to submillimeter emission from the HD~109573A disk.

\paragraph{HD 102647 ($\beta$ Leo).}

Our disk parameter estimates for HD~102647 ($T_{\rm
dust}=100\pm10$~K, $M_{\rm dust}\gtrsim1.7\times10^{-3}M_\earth$) 
agree with those found in the literature.  \citet{lau02a} derive a dust 
temperature of
$83\pm5$~K for HD~102647 from the 25$\micron$/60$\micron$ {\sl ISO} flux 
ratio, while \citet{jay01} find $T_{\rm dust}=120$~K from blackbody fits to
the photosphere and to the IR excess in this star.  The amount of dust 
around HD~102647 inferred from the observations of \citeauthor{lau02a} is 
$1.6\times10^{-3}M_\earth$:
a number not given in their paper, but calculated here based on their
assumptions (greybody particles) and their formula (A.5).  From 
submillimeter upper limits, on the other hand, \citet{hol03} find
$M_{\rm dust}<0.59M_{\rm Moon}=7.3\times10^{-3}M_\earth$
\citep[$M_{\rm Moon}=M_\earth/81.301$;][]{hei91}.  Both results are 
consistent with the one presented here.

\paragraph{HD 17925.}

\citet{hab01} find evidence for a significant {\sl ISO} excess in 
HD~17925 only at 60$\micron$ (with an upper limit on the flux at 90$\micron$),
and, assuming a disk distance of 50~AU ($T_{\rm eff}\approx 30$~K) from 
the star (similar to that in the Vega system), find
$M_{\rm dust}=1.08\times10^{-3}M_\earth$: a value which agrees with our 
range of estimates.  However, by
considering the full range of blackbody temperatures consistent with the IR
data, we obtain a range of possible masses of the dust disk.  Indeed,
our measured 1.3-sigma $N$ band excess in HD~17925, combined with
a 1.3-sigma excess at 12$\micron$ ($\Delta[12]=0.09\pm0.07$, 
from {\sl IRAS}), 
and a 2.7-sigma excess at 25$\micron$ ($\Delta[25]=0.27\pm0.10$ from
the averaged {\sl ISO} and {\sl IRAS} flux densities, as adopted by
\citealt{lau02a}), point to the possible presence of warmer (140~K)
material orbiting the star.  
Recent mid-IR spectroscopic observations by \citet{gai03}, also point to
the marginal presence of a silicate emission peak at 7\% of the
photosphere at 10$\micron$.

The IR SED of this star can be fit by a 
blackbody as warm as $\sim$250~K (Figure~2), producing an excess of 
$f_d\approx3\times10^{-4}$.  The debris would orbit
at a distance of $\sim7$~AU if the particles are greybodies ($\sim$0.8~AU if
blackbodies), and would have 
a mass $M_{\rm dust}\geq7\times10^{-6}M_\earth$ (lower end of the range listed
in Table~\ref{tab_disks}), which could make this 
the least massive debris disk detected outside our Solar System.  The required
mass of parent bodies would be $M_{\rm PB}\geq0.015M_\earth$, or $\gtrsim$3
times the mass of the main asteroid belt in the Solar System \citep{bin00}.  

Unfortunately, at the level of the 10--25$\micron$ excesses seen in HD~17925 
(1--2$\sigma$), we are strongly affected by the accuracy
of our assumptions for the model photosphere (mainly the stellar metallicity),
as well as by the precision of the $JHK$ photometry for the object, 
used for the normalization of the model photosphere.  Whereas the level of 
the 10--25$\micron$ excess is 2.0--2.5$\sigma$ if the NextGen photosphere is
normalized to the $J$ or $H$ apparent magnitude, its significance decreases to
0.3--2.0$\sigma$ if the normalization is done at $K$.  

A modest amount of extinction ($A_V\approx0.2$~mag), a sub-solar metallicity 
(${\rm [Fe/H]}\lesssim-1.5$), a sub-dwarf gravity, or a cooler photosphere 
(by $\sim200$~K) could all
account for the higher mid-IR fluxes in
HD~17925.  However, given $f_d\leq2\times10^{-4}$ (Table~\ref{tab_excesses})
and the vicinity of the star to the Sun ($10\pm0.1$~pc) we expect $A_V<0.01$.
The metal abundance and gravity of the star is very nearly solar 
\citep{hay01,cay89,per83},
and its temperature is estimated from high-resolution spectroscopy at 5091~K
\citep{cay89,per83}.  None of these factors are thus expected to change
the expected stellar SED of HD~17925 between 10--25$\micron$.

Higher-precision ($<0.10$~mag) mid-IR photometry could help constrain the 
temperature and
extent of the debris disk around HD~17925, but would be difficult to obtain.
However, submillimeter imaging can further constrain the SED of the star,
and could potentially even resolve the emission around it, given its
vicinity (10~pc).

\subsubsection{Newly-reported Debris Disks: HD~107146 and 
HD~88638 \label{sec_88638}}

\paragraph{HD 107146.}

From our analysis of the {\sc IRAS} FSC data on HD~107146, we report a new
excess around this star 
(although, see \citet{wil03} for an independent discussion of
a newly-observed sub-millimeter excess in this source) 
at wavelengths $\geq60\micron$, likely due
to an orbiting debris disk.  Despite a relatively high dust luminosity
fraction, $f_d=1.5\times10^{-3}$ (the {\sl IRAS} excesses from known 
Vega-like
stars range from $10^{-3}-10^{-5}$), it has remained overlooked due to its
apparent faintness.  The inferred minimum dust mass $M_{\rm dust}$ 
(for GB dust particles) is comparable to that in HD~109573A,
and should be detectable in the sub-millimeter, or in visible scattered light 
observations.   These can then be used to further constrain the spatial 
extent, morphology, and mass of the debris disk.  HD~107146 is one
of very few main-sequence G dwarfs known to harbor debris 
disks \citep[see][for other candidates in addition to 
$\epsilon$~Eri]{man98,syl00,dec00}.  
At its intermediate age (50--300~Myr; Table~\ref{tab_phot}) it 
helps bridge the gap between the epoch of formation of the Solar
System, and its current evolved state of low dust content.

\paragraph{HD 88638.}

The evidence for a debris disk around 
HD~88638 is only tentative, expressed as a single 2.4-sigma excess at
10.3$\micron$.  The photosphere of the star is below the sensitivity limits
of the {\sl IRAS} All-Sky Survey, and it has not been targeted with {\sl ISO}.
Longer-wavelength, sensitive IR and sub-millimeter observations (e.g., with
{\sl SIRTF}) are needed to 
confirm the existence of a disk around this star.  At the age
of HD~88638 ($\sim$600~Myr, Table~\ref{tab_phot}), a debris disk of 
the inferred mass and distance from the star (Table~\ref{tab_disks}) would bear
the closest resemblance to the main asteroid belt in our Solar System among
all other known stars with debris disks.  

In fitting the excess in the SED of HD~88638, we observe the same 
trend as discussed in the case of HD~17925
(Section~\ref{sec_diskparams}).  Because the temperature of the
blackbody required to fit the excess is not well-constrained, the disk 
luminosity fraction $f_d$ and the required dust mass $M_{\rm dust}$ are 
not correlated, and are temperature-dependent.

\subsection{Dust Mass Limits for Stars Not Detected in Excess}

With the exception of HD~147809, all the stars in our sample that lack an 
excesses are of spectral type G0 and
later.  The derived upper limits for $f_d$ (assuming 300~K debris disks)
indicate $M_{\rm dust}\lesssim10^{-5} M_\earth$, assuming that the
dust particles behave like greybodies.  For comparison, the minimum dust
mass detectable around a G0 dwarf in the {\sl ISO} 25$\micron$ survey of
\citet{lau02a} is estimated at $2\times10^{-6}M_\earth$, assuming a 120~K
disk.  Our survey appears overall slightly less sensitive than theirs,
possibly because of our slightly larger
median errorbars: $\pm0.11$~mag in $\Delta N$ for our ground-based 
observations vs.\
$\pm0.10$~mag in $\Delta [25]$ for the {\sl ISO} observations of
\citet{lau02a}.  However, if we apply the method of dust mass estimation
of \citeauthor{lau02a}, we obtain that our $N$ band observations should be
sensitive to disk masses $M_{\rm dust}\gtrsim1\times10^{-6}M_\earth$ 
around G0 stars, 
i.e.\ our sensitivity is comparable to theirs in terms of the minimum amount
of detectable dust mass.  Regardless, from the standpoint of searching for 
debris disks of different temperatures (and thus, radii), the two 
surveys are complementary.

\subsection{Fractional Excess and Debris Disk Mass vs.\ Stellar Age
\label{sec_fdmd}}

The results of two large studies targeting disk evolution around 
main-sequence stars 
with {\sl ISO} and {\sl IRAS} data have shown that the amount of 
circumstellar dust,
and the number of stars possessing detectable IR dust excesses, are both
decreasing functions of stellar age 
\citep[][see also \citealt{hab01}]{spa01,hab99}.
Given our very limited sample, and a bias in the source selection toward 
stars with known dust disks we cannot test the
conclusions of \citeauthor{hab99} and \citeauthor{spa01}  We do, 
however, reproduce the tendency
of decreasing dust fraction $f_d$ with age among stars with detected IR
excesses (Figure~\ref{fig_agefdmd}; upper panel),
with a power law index of $-2.0\pm0.5$ ($\chi^2_{\rm
p.d.f.}=1.6$), compared with --1.76 derived in \citet{spa01}.  
AA~Tau and HD~88638 have been excluded from the fit, 
the dust disk of AA~Tau being optically thick (and hence not relevant to
the discussion of debris disk evolution), and the excess in
HD~88638 being uncertain.  Because we are biased towards detecting 
only the most luminous debris disks, our estimates of $f_d$ 
delineate only an upper envelope to the actual amount of IR excess
present in any object in the 1~Myr~-- 1~Gyr age range.

As discussed in Section~\ref{sec_diskparams} above,
the amount of IR excess $f_d$ is not in a one-to-one correspondence with
the mass of emitting circumstellar dust $M_{\rm dust}$, but is also 
a function of the dust temperature, or equivalently, of the distance $D$ 
of the dust from the star (Equation~3).
Hotter dust produces larger values of $f_d$ by virtue of a
larger area under the Planckian curve, for a fixed dust mass and mean
particle size.  It would therefore be a coincidence 
if the $f_d$-age and $M_{\rm dust}$-age relations followed the same 
functional form.
This would require that circumstellar dust is maintained at the same
temperature around each star, whereas we have examples from our own Solar
System (zodiacal dust vs.\ asteroid belt vs.\ Kuiper belt), and from other
Vega-like systems (e.g., $\zeta$~Lep vs.\ $\alpha$~Lyr vs.\ $\epsilon$~Eri)
that dust temperatures range from $\sim30-260$~K.  One could therefore
expect a tighter correlation in the $\log(M_{\rm dust})-\log({\rm age})$
vs.\ the $\log(f_d)-\log({\rm age})$ relation, given that the former is
a more physical relationship.

A plot of $M_{\rm dust}$ as a function of stellar
age is presented in the bottom panel of Figure~\ref{fig_agefdmd}.  The
fitted slope is $\log(M_{\rm dust}) \propto (-3.5\pm0.9)\times \log
({\rm age})$
(excluding AA~Tau and HD~88638), with a goodness of fit $\chi^2_{\rm
p.d.f.}=1.7$, similar to the one found for the $f_d$-age relation above.  
There is thus no
evidence from our limited amount of data that $M_{\rm dust}$ is more
tightly correlated with stellar age than $f_d$.  The $M_{\rm dust}$-age
relation is however steeper than the one
derived by \citet{spa01}, and marginally (1.7$\sigma$) inconsistent with that
expected for collisionally replenished secondary dusts disks
\citep[power-law index near --2;][]{zuc93}.  The inclusion of the data
point due to HD~88638 only strengthens this trend ($\chi^2_{\rm p.d.f.}=1.3$
vs.\ $\chi^2_{\rm p.d.f.}=1.7$ for the $M_{\rm dust}$-age vs.\ $f_d$-age
case), with the power-law indices remaining
approximately unchanged ($-3.6\pm0.7$ and $-1.7\pm0.4$, respectively), 
though slightly more discrepant with each other (at the 2.4$\sigma$ 
level, vs.\ 1.5$\sigma$ when HD~88638 is excluded).

Given that $M_{\rm dust}$ is a direct estimate of the dust disk mass,
and the marginal inconsistency between the two power law indices, it
appears that $f_d$ may be a less adequate tracer
of the evolution of dust disks than our estimate of $M_{\rm dust}$.
The 

\begin{equation}
M(\mbox{emitting dust})~\propto f_d \propto \frac{dN}{dt} \propto 
	t_{\rm age}^{-2}
\end{equation}

\noindent relation is inferred by \citet{spa01} based on the purely geometric
assumption that the collisional rate of large particles $dN/dt$ goes
as $N^2$, where $N$ is the number density of large planetesimals in
the system.  When the gravitational potentials $G m/a$ of the
planetesimals become comparable to the square of their random velocities
$v^2_{\rm rms}$, the collisional cross-sections of the planetesimals 
increases beyond their
simple geometric cross-sections because of gravitational focusing.  
In the two-body approximation, the collision rate is given by
\citep[e.g.,][]{saf69}:

\begin{equation}
\sigma_{\rm g} v_{\rm rms} = \pi a^{2}
	\left (1+\frac{2 G m}{a v_{\rm rms}^{2}} \right ) v_{\rm rms},
\end{equation}

\noindent where $\sigma_{\rm g}$ is the gravitational cross-section.
Collisions between planetesimals become more likely with increasing
particle size, and their rate goes as $a^4$,
instead of $a^2$ as in the geometrical regime.  The resulting
runaway and/or oligarchic growth
\citep{gre78,kok95} should effectively increase the depletion rate of
dust particles faster than $dN/dt \propto t_{\rm age}^{-2}$; i.e.,

\begin{equation}
M(\mbox{emitting dust})~\propto \frac{dN}{dt} \propto t_{\rm age}^{-x}, 
\end{equation}

\noindent where $x>2$.  Hence, the steeper power law index that we infer 
($x \sim -3.5$) presents tentative evidence for planetesimal growth
in debris disks.

An important shortcoming of the above discussion is that we do not consider 
the stars for which we do not find evidence for orbiting debris disks.
As noted, due to an inherent bias toward stars with known
debris disks in our sample, we cannot perform an analysis similar to the 
ones in \citet{hab01} and \citet{spa01}.  It is significant however, that
a number of our sources do not show any IR excess, when they would be
expected to possess such, given universal application of the inferred 
$\log(M_{\rm dust}) - \log({\rm
age})$ relation.  The apparent scatter in disk mass vs.\ age may depend on 
the history of 
each system, and in particular, on whether favorable conditions for disk
retainment had been present throughout the evolution of the star.  For
instance, the youngest star in our sample, DI~Tau, shows no excess emission at
10$\micron$, a rather atypical state for an M star $\lesssim1$~Myr old.
DI~Tau is however a binary system with a $\sim20$~AU projected separation, and
\citet{mey97} argue that the formation of the companion has led to the rapid
evolution of the circumstellar disk that may have surrounded DI~Tau.  On the
other hand, HD~145519 \citep{mat89} and HD~109573A also
reside in binary systems, though exhibit IR excesses.  The separation
between the components, or conditions other than binarity may be 
important for the disk dissipation time 
scale.  The spectral type of the star may play a role \citep{dec00}, 
with hotter stars
eroding their disks faster, as a result of higher radiation pressures, and
shorter Poynting-Robertson time scales.  In addition, the presence or
absence of orbiting planets may also affect disk-clearing \citep{gol73}.

A larger, unbiased sample, including a range of stellar spectral types, as
well as single and multiple systems would be needed to address the issues
raised here.  More accurate dust mass estimates from disk models,
employing realistic grains, would be 
needed to verify the validity of our suggestion
that $M_{\rm dust}$ (as calculated from Equation~3) is a steeper 
function of stellar age than $f_d$, and potentially a more adequate
tracer of the evolution of debris disks.

\section{CONCLUSION}

Using ground-based 10$\micron$ observations we have demonstrated sensitivity
to the stellar photosphere of nearby solar-type stars, and hence to small
excess ($f_d=L_{\rm IR}/L_* \approx10^{-3}$ for a 300~K debris disk) above 
it.  Our mid-IR
data confirm previous space- and ground-based observations for a subset of our
targets, and give new insight into this regime for the remainder, by
complementing the existing data.  By employing simple fitting of the observed 
IR excesses with blackbody curves, we extract reliable estimates of the ratio
$f_d$ of the dust luminosity to the stellar bolometric luminosity.  We
calculate lower limits on the amount of dust present in each system, after
accounting for the effects of radiation pressure and Poynting-Robertson drag.
Our analysis of the excess in the HD~17925 system points to the fact that it
may be the star with the smallest amount of circumstellar dust detected among
Vega-like systems to date.

We present strong evidence for the existence of a previously undetected 
debris disk around HD~107146, and possibly one around HD~88638.
Our detection of an excess in HD~107146 from
archival {\sl IRAS} FSC data suggests that there may be more stars with
unknown debris disks in the deeper and higher signal-to-noise FSC, than 
what is 
currently known from analyses of the PSC.  A methodical search for these in
combination with 2MASS near-IR photometry, as already presented in
\citet{faj00} for the PSC, but watchful of the caveats addressed here
and the ones described in 
\citet{son02}, could therefore prove fruitful.  Given that very few disks
are found around late-type stars due to their intrinsic faintness,
the results of such a search could significantly increase the number of known 
debris disks around solar- and later-type stars.  Ground-based 10$\micron$ 
and 20$\micron$ follow-up observations will be indispensable in discerning 
{\sl IRAS/ISO} false positives from
bona-fide mid-IR excesses, as well as potentially successful in finding new 
excess candidates (e.g., as in the case of HD~88638).  

Based on our limited data, we discuss the possibility of breaking the
degeneracy between debris disk temperature and mass in the $f_d$ vs.\
stellar age relation.  We find that the power law index in the inferred
dust mass $M_{\rm dust}$-age relation is steeper than in 
the $f_d$-age case, and hence tentatively propose the use of 
$M_{\rm dust}$-age relation as a better tracer of the evolution of the 
amount of debris around
main-sequence stars.  However, a deficiency of the proposed power-law is 
that
it does not take into account the absence of debris disk around stars which 
would
be expected to harbor such based on their age.  We argue that other factors,
such as stellar multiplicity and spectral type, may also play a role.

Future space-based IR observations with {\sl SIRTF} and {\sl SOFIA} are 
expected to
greatly expand the number of known Vega-like stars, by being sensitive
to debris disks as faint as the zodiacal dust 
\citep[$f_d\sim8\times10^{-8}$][]{bac93} around nearby
stars as evolved as our Sun.  These will provide the database necessary to
answer the prominent questions about 
the processes and time-scales of evolution of debris disks around
main-sequence stars, and will place our own solar system in the context of
stellar and planetary evolution.

\begin{acknowledgements}

We would like to thank Randy Campbell and Gregory Wirth for their
assistance with LWS, Tom Hayward and Rick Burress for help with SC--10,
and Jonathan Foster for help with the LWS data reduction.
This research has made use of the IPAC InfraRed Science
Archive, which is operated by the California Institute of Technology, under 
contract with the NASA, and of the
SIMBAD database, operated at CDS, Strasbourg, France.
Data presented in this paper were analyzed at ESA's ISO Data
Centre at Vilspa, Spain.
The publication makes use of data products from the Two Micron All Sky
Survey, which is a joint project of the University of Massachusetts and
the IPAC/California Institute of Technology, funded by the NASA and the NSF.

\end{acknowledgements}

\clearpage

\begin{figure}
\figurenum{1a}
\plotone{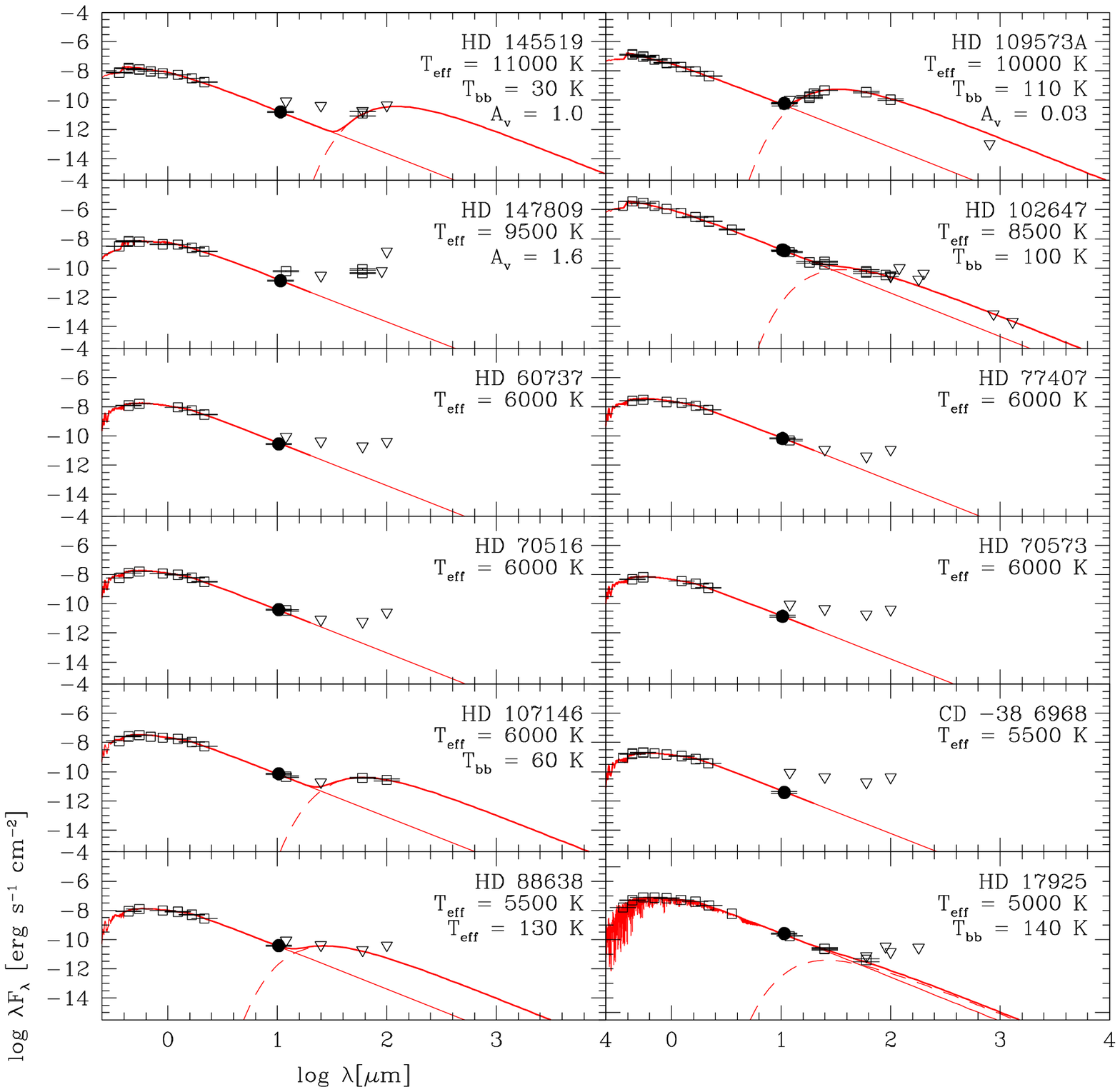}
\figcaption{Spectral energy distributions of the observed stars.  Our
10$\micron$ data are plotted with solid circles, while data
from the literature are plotted with open squares (with errorbars 
where available).  Upper limits are plotted as downward
facing triangles.  Photospheric models (thin lines;
photospheric temperatures listed under the object names) are 
from \citet{kur79,kur92} and from \citet{hau99}.  Dashed lines represent
blackbody fits to the observed excess emission. 
The total flux densities (photosphere + blackbody) are plotted with thick 
lines.}
\end{figure}

\begin{figure}
\figurenum{1b}
\plotone{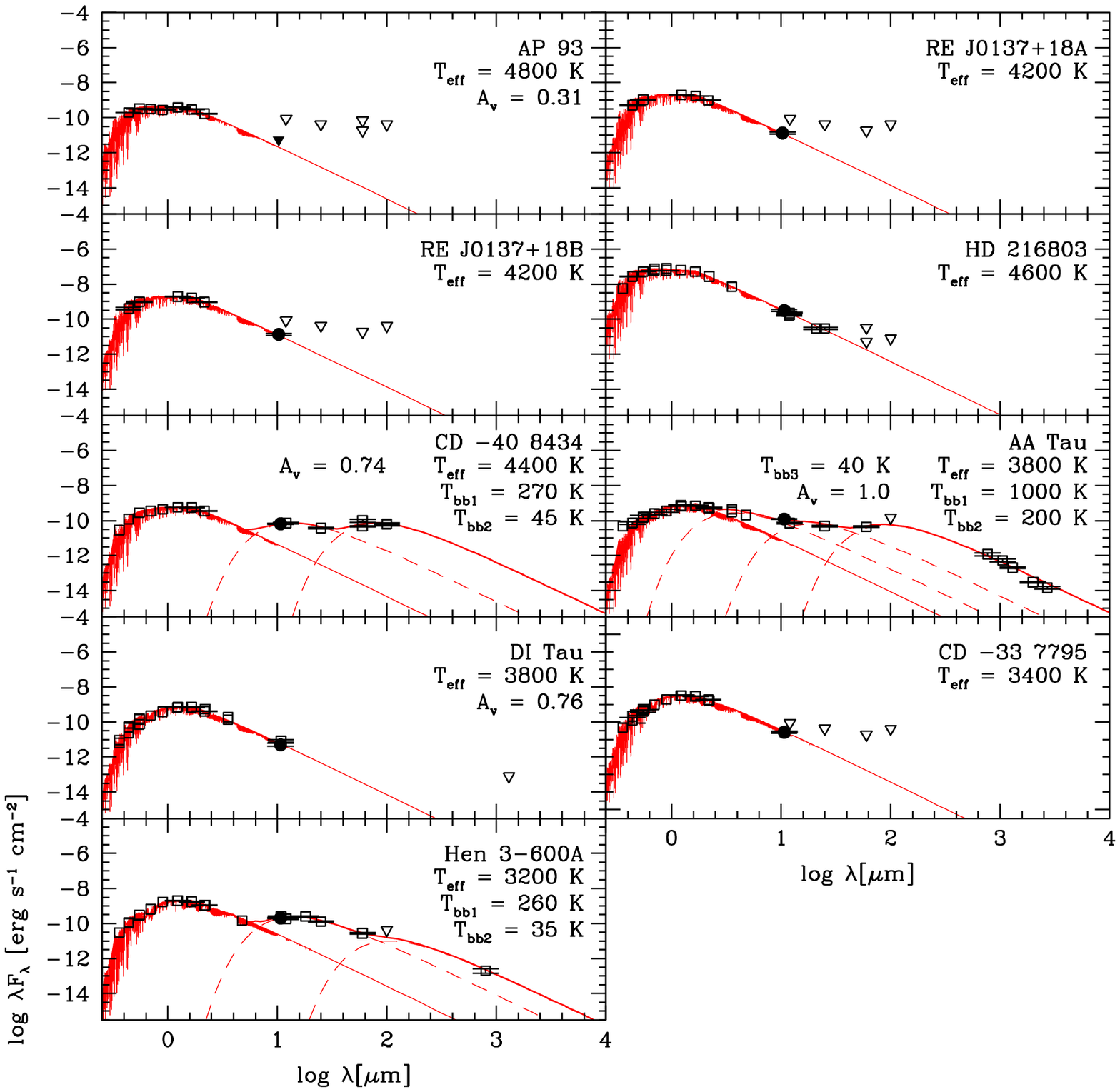}
\figcaption{
}
\end{figure}

\begin{figure}
\figurenum{2}
\plotone{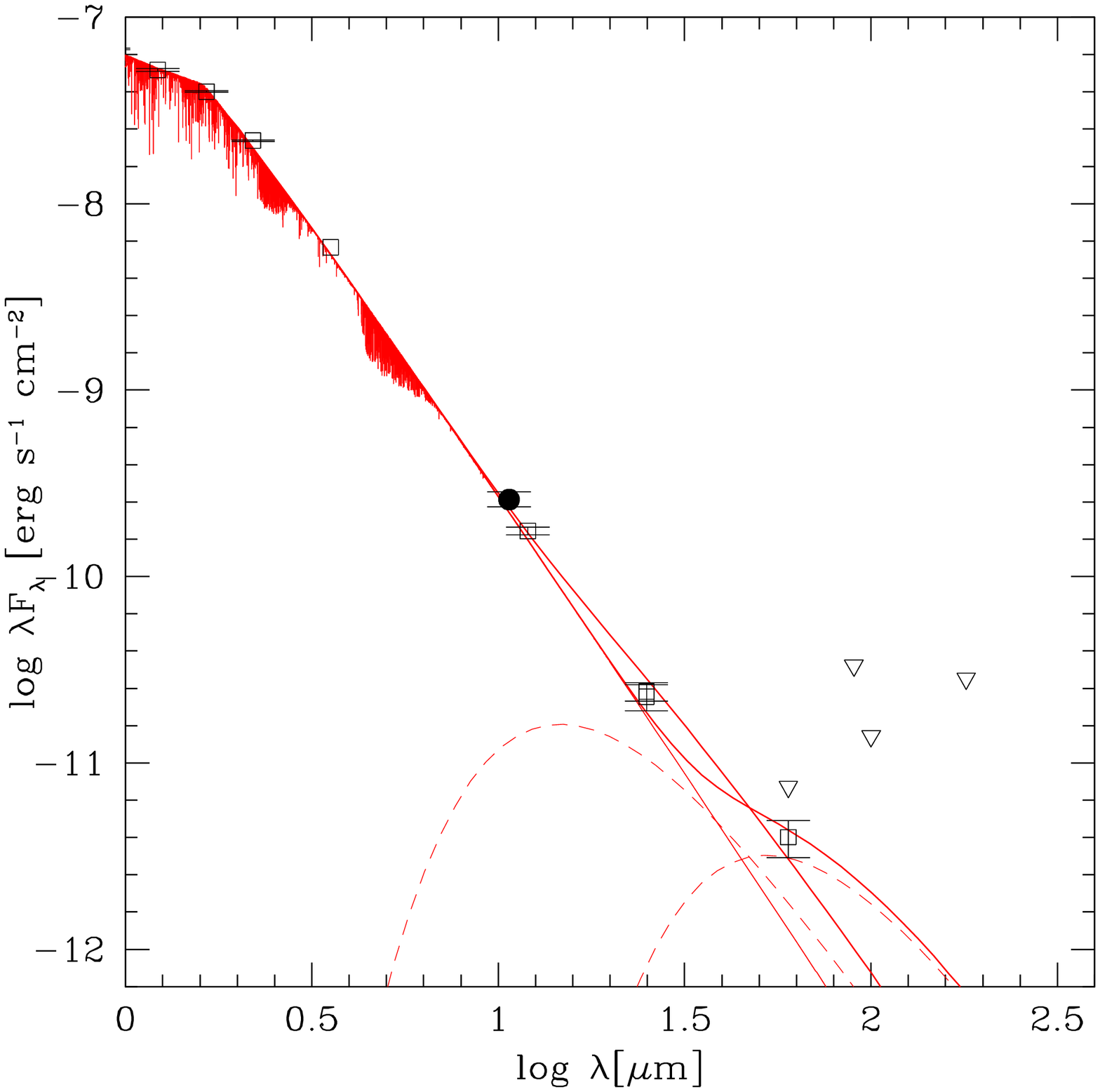}
\figcaption{Fits to the SED of HD 17925, where 250~K and 70~K
blackbodies have been allowed to independently account for the excess.
Symbols are as in Figure~1.  The NextGen photosphere has
been normalized to the $J, H$, and $K$ band measurements, and
the temperature and normalization of the blackbody fit to the excess
have been allowed to vary freely.  The two cumulative curves (thick
continuous lines) fit the 10--60$\micron$ data comparatively well
($\chi^2_{\rm p.d.f.}=2.3$ and 2.5, respectively).  However, the hotter
(250~K) blackbody produces a higher fractional excess $f_d$ than the cooler
(70~K) one by half an order of magnitude, whereas the amount of dust 
$M_{\rm dust}$ required is less by nearly an order of magnitude.  The
best fit is achieved with a 140~K blackbody, corresponding to $M_{\rm
dust}\approx 2\times10^{-6} M_\earth$.
\label{fig_17925}}
\end{figure}

\begin{figure}
\figurenum{3}
\plotone{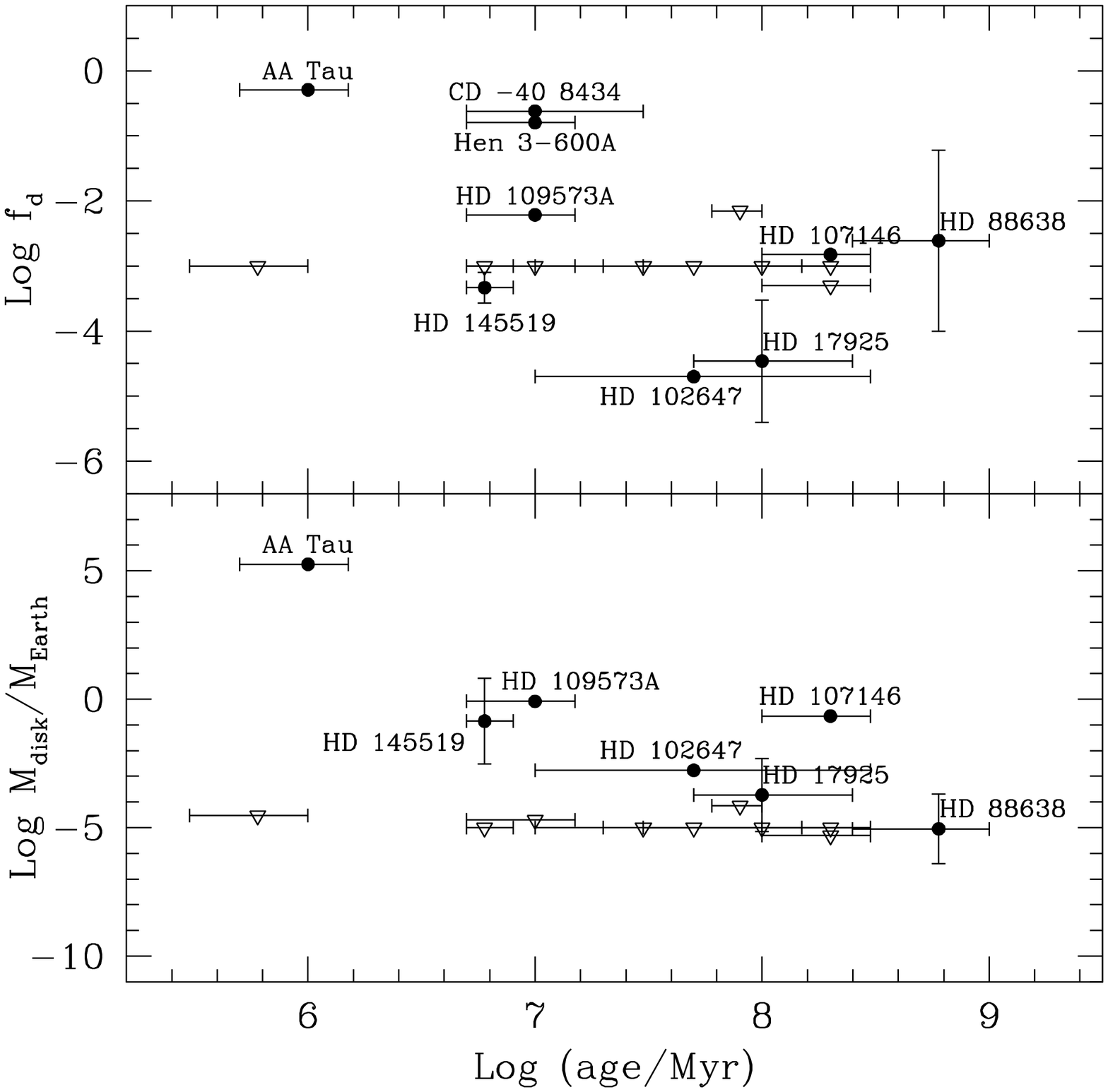}
\figcaption{{\sl Top panel:} $f_d$ vs.\ stellar age plot. {\sl Bottom panel:} 
$M_{\rm disk}$ vs.\ stellar age plot.  In both panels solid dots represent
stars with detected IR excesses, and are labeled.  Down-pointing open 
triangles represent upper limits on $f_d$, or on the dust mass $M_{\rm
dust}$, in systems with no excess IR emission, and are not labeled.  
Vertical errorbars are given for HD~17925 and HD~88638,
corresponding to the range of inferred dust masses.  The errorbars on the
other points are $\sim0.5$~dex in the top panel, and $\sim1.0$~dex in
the bottom panel.  The data are from Tables~\ref{tab_excesses}
and \ref{tab_disks}, assuming greybody (GB) particle properties.
\label{fig_agefdmd}}
\end{figure}

\clearpage

\begin{deluxetable}{llccl}
\tablecaption{Observational Epochs \label{tab_dates}}
\tablehead{ \colhead{Star} & \colhead{Other ID} & \colhead{Date (UT)} & 
	\colhead{Instrument} & \colhead{Association}}
\startdata
AA Tau & IRAS F04318+2422 & 2000 Feb 21 & LWS & Taurus CTTs\\
DI Tau & & 2000 Feb 21 & LWS & Taurus WTTs\\
CD $-$33$\degr$7795 & TWA 5A & 2000 Feb 20 & LWS & TW Hya WTTs\\
Hen 3$-$600A & TWA 3A & 2000 Feb 20 & LWS & TW Hya CTTs\\
HD 109573A & HR 4796A & 2000 Feb 21 & LWS & TW Hya\\
CD $-$38$\degr$6968 & RX J1109.7$-$3907 & 2000 Dec 9 & LWS & 
	TW Hya WTTs\tablenotemark{\dag} \\
HD 147809 & HIP 80425 & 2000 Feb 20 & LWS & Upper Sco\\
HD 145519 & SAO 159767 & 2000 Feb 20 & LWS & Upper Sco\\
HD 102647 & HIP 57632 & 2000 Feb 20 & LWS & field star\\
HD 17925 & HIP 13402 & 2000 Feb 21 & LWS & field star\\
CD $-$40$\degr$8434 & IRAS 14050$-$4109 & 2000 Feb 21 & LWS & field WTTs\\
HD 216803 & HIP 113283 & 2000 Dec 9 & LWS & field star\\
AP 93 & V525 Per & 2002 Jan 1 & SC--10 & $\alpha$ Per\\
RE J0137+18A & RX J0137.6+1835A & 2002 Jan 1 & SC--10 & field star\\
RE J0137+18B & RX J0137.6+1835B & 2002 Jan 1 & SC--10 & field star\\
HD 60737 & HIP 37170 & 2002 Jan 1 & SC--10 & field star\\
HD 70573 & SAO 116694 & 2002 Jan 3 & SC--10 & field star\\
HD 70516 & HIP 41184 & 2002 Jan 3 & SC--10 & field star\\
HD 77407 & HIP 44458 & 2002 Jan 3 & SC--10 & field star\\
HD 88638 & HIP 50180 & 2002 Jan 3 & SC--10 & field star\\
HD 107146 & HIP 60074 & 2002 Jan 3 & SC--10 & field star\\
\enddata
\tablenotetext{\dag}{Candidate TWA member \citep{ste99}.}
\end{deluxetable}

\clearpage

\begin{deluxetable}{lcccccccccc}
\tabletypesize{\scriptsize}
\rotate
\tablewidth{0pt}
\tablecaption{Infrared photometry and properties of the observed sources
	\label{tab_phot}}
\tablehead{\colhead{Source} & \colhead{$J-H$} & \colhead{$H-K_s$} & 
	\colhead{$K_s-N$} & \colhead{$N$} & \colhead{$\Delta N$} &
	\colhead{$K_s-$[12]} & \colhead{sp.\ type} & \colhead{age} & 
	\colhead{parallax} & Notes\tablenotemark{\dag}\\
	 & \colhead{(mag)} & \colhead{(mag)} & \colhead{(mag)} & 
	 \colhead{(mag)} & \colhead{(mag)} & \colhead{(mag)} & & 
	 \colhead{(Myr)} & \colhead{(mas)} & }
\startdata
HD 145519 & $0.18\pm0.04$ & $0.14\pm0.05$ & $-0.44\pm0.10$ & $7.04\pm0.10$ 
	& $-0.10\pm0.13$ & $<1.29$ & B9V & 5--8 & $6.3\pm0.8$ & 1,2,3\\
HD 109573A & $0.00\pm0.08$ & $0.00\pm0.08$ & $0.22\pm0.12$ & $5.58\pm0.10$ & 
	$0.29\pm0.11$ & $<1.25$ & A0V & $10\pm5$ & $14.91\pm0.75$ & 1,4\\
HD 147809 & $0.22\pm0.06$ & $0.16\pm0.06$ & $-0.15\pm0.12$ & $7.21\pm0.12$ & 
	$-0.11\pm0.13$ & $1.91\pm0.11$ & A1V & 5--8 & $6.3\pm0.8$ & 2,3\\
HD 102647 & $0.34\pm0.02$ & $0.01\pm0.01$ & $-0.08\pm0.08$ & $2.06\pm0.08$ & 
	$-0.03\pm0.09$ & $0.03\pm0.06$ & A3V & 10--300 & $90.16\pm0.89$ & 
	5,6 \\
CD $-$38$\degr$6968 & $0.29\pm0.06$ & $0.03\pm0.05$ & $-0.12\pm0.19$ & 
	$8.60\pm0.19$ & $-0.10\pm0.19$ & $<2.96$ & G3 & $<100$ & 
	$7.90\pm23.80$ & 7 \\
HD 107146 & $0.27\pm0.03$ & $0.07\pm0.03$ & $0.19\pm0.09$ & $5.35\pm0.09$ & 
	$0.11\pm0.09$ & $0.05\pm0.13$ & G2V & 50--250 & $35.07\pm0.88$ & 
	8,9 \\
HD 60737 & $0.28\pm0.03$ & $0.06\pm0.03$ & $-0.11\pm0.11$ & $6.36\pm0.11$ & 
	$-0.18\pm0.11$ & $<0.73$ & G0 & 50--250 & $26.13\pm1.04$ & 
	9,10 \\
HD 70573 & $0.28\pm0.04$ & $0.09\pm0.04$ & $0.05\pm0.18$ & $7.14\pm0.18$ & 
	$-0.01\pm0.18$ & $<1.67$ & G1/2V & 50--250 & $11.30\pm10.70$ & 
	9,11 \\
HD 70516 & $0.28\pm0.03$ & $0.09\pm0.02$ & $0.14\pm0.09$ & $6.00\pm0.09$ & 
	$0.07\pm0.09$ & $0.39\pm0.11$ & G0 & 100--300 & $27.10\pm2.18$ & 
	9 \\
HD 77407 & $0.27\pm0.03$ & $0.09\pm0.03$ & $-0.01\pm0.08$ & $5.45\pm0.08$ & 
	$-0.08\pm0.08$ & $0.07\pm0.11$ & G0 & 20--150 & $33.24\pm0.91$ & 
	9,12 \\
HD 88638 & $0.27\pm0.03$ & $0.11\pm0.03$ & $0.29\pm0.07$ & $6.04\pm0.07$ & 
	$0.19\pm0.08$ & $<0.81$ & G5 & 250--1000 & $26.66\pm3.09$ & 9 \\
HD 17925 & $0.39\pm0.02$ & $0.09\pm0.01$ & $0.02\pm0.10$ & $4.00\pm0.10$ & 
	$0.14\pm0.11$ & $0.01\pm0.05$ & K1V & 50--250 & $96.33\pm0.77$ & 
	5,8,9 \\
HD 216803 & $0.54\pm0.02$ & $0.11\pm0.01$ & $0.03\pm0.14$ & $3.80\pm0.14$ & 
	$0.06\pm0.14$ & $0.07\pm0.06$ & K4V & 200$\pm$100 & $130.94\pm0.92$ & 
	5,9,13 \\
CD $-$40$\degr$8434 & $0.73\pm0.05$ & $0.28\pm0.05$ & $3.05\pm0.10$ & 
	$5.49\pm0.10$ & $3.14\pm0.12$ & $3.56\pm0.09$ & K5 & $\lesssim$10 & 
	\nodata & 14 \\
RE J0137+18A & $0.62\pm0.03$ & $0.14\pm0.03$ & $0.26\pm0.14$ & 
	$7.18\pm0.14$ & $0.15\pm0.15$ & $<1.92$ & K3V & 30--55 & 
	$15.8\pm5.0$ & 9,12,15,16,17 \\
RE J0137+18B & $0.61\pm0.03$ & $0.15\pm0.03$ & $0.32\pm0.14$ & 
	$7.23\pm0.14$ & $0.22\pm0.15$ & $<1.91$ & K3V & 30--55 & 
	$15.8\pm5.0$ & 9,12,15,16,17 \\
AP 93 & $0.50\pm0.03$ & $0.14\pm0.03$ & $<1.18$ & $>8.18$ & 
	$<1.1$ & $<3.84$ & K2 & 80 & 5.3 & 18,19 \\
AA Tau & $0.87\pm0.05$& $0.51\pm0.05$ & $3.23\pm0.08$ & $4.81\pm0.07$ & 
	$3.07\pm0.09$ & $3.58\pm0.11$ & M0V & 1 & 7.1 & 20\\
DI Tau & $0.72\pm0.03$ & $0.21\pm0.03$ & $0.10\pm0.22$ & $8.29\pm0.22$ & 
	$-0.08\pm0.22$ & \nodata & M0.5V & 0.6 & 7.1 & 21 \\
CD $-$33$\degr$7795 & $0.68\pm0.04$ & $0.24\pm0.04$ & $0.24\pm0.10$ & 
	$6.50\pm0.10$ & $-0.11\pm0.10$ & $<1.23$ & M3V & $10\pm5$ & 
	$18\pm3$ & 4,22 \\
Hen 3$-$600A & 0.62 & 0.30 & $3.04\pm0.20$ & $4.26\pm0.19$ & 
	$2.54\pm0.10$ & $3.48\pm0.09$ & M4 & $10\pm5$ & $18\pm3$ & 
	4,22,23,24 \\
\enddata
\tablenotetext{\dag}{1.\ $JHK$ photometry from \citet{jur93}; 
2.\ Assumed age for Upper Sco OB association is 5--8~Myr \citep{deg89};
3.\ Mean distance for Upper Sco from \citet{deg89,jon70};
4.\ Assumed TW~Hya age is $10\pm5$~Myr; 
5.\ $JHK$ photometry from \citet{aum91}; 
6.\ Maximum age estimate from \citet{son01}.  Minimum age estimated
  from the lack of known accretion signature in this star; 
7.\ Age estimate from \citet{ste99};
8.\ Age deduced from Li measurement of \citet{wic03}; 
9.\ Age estimate from comparison of unpublished Li~$\lambda$6707 
  equivalent width to open cluster sequence;
10.\ Spectral type from \citet{hou99}; 
11.\ Age deduced from Li measurement reported in \citet{jef95};
12.\ Age estimate from \citet{mon01}; 
13.\ Age estimate from \citet{bar98}; 
14.\ Age estimate based on Li measurement and spectral type estimate
  in \citet{gre02}; 
15.\ The binary is unresolved in 2MASS.  $JHK_s$ photometry from adaptive 
  optics observations by Metchev \& Hillenbrand, in prep.;
16.\ Spectral type from \citet{jef95}; 
17.\ Distance estimate from \citet{mon01}; consistent with being a
  nearby member of the Cas-Tau OB association (E. Mamajek 2003, private
  communication);
18.\ 3$\sigma$ upper limit at 10$\micron$;
19.\ Spectral type deduced from effective temperature listed in 
  \citet{ran98}; 
20.\ Age estimate from \citet{bec90};
21.\ Age estimate from \citet{mey97}; 
22.\ Distance obtained as the average of the distances to the 4
  TWA members with Hipparcos astrometry: TW~Hya, HR~4796A, HD~98800, and
  TWA~9 (56, 67, 47, and 50 pc, respectively); 
23.\ The binary is unresolved in 2MASS.  $JHK$ photometry from \citet{web99};
24.\ Spectral type from \citet{tor00}.}
\end{deluxetable}

\clearpage

\begin{deluxetable}{lcccccc}
\tabletypesize{\scriptsize}
\tablewidth{0pt}
\tablecaption{Fractional disk excesses \label{tab_excesses}}
\tablehead{\colhead{Star} & \colhead{Age [Myr]} & 
	\colhead{$T_{\rm dust}$ [K]} & \colhead{$f_d$}
	& \colhead{$f_{d {\rm (Lit)}}$} & 
	\colhead{Reference} & Excess?}
\startdata
HD 145519 & $5-8$ & $30-140$ & $2.7-8.0\times10^{-4}$ & \nodata & 1 & Yes\\
HD 109573A & $10\pm5$ & $110\pm10$ & $6.1\times10^{-3}$ & 
	$5-10\times10^{-3}$ & 2,3 & Yes\\
HD 147809 & $5-8$ & (300)\tablenotemark{\dag} & $<5\times10^{-4}$ & 
	\nodata & \nodata & No \\
HD 102647 & $10-300$ & $100\pm10$ & $2.0\times10^{-5}$ & 
	$1.2-1.9\times10^{-5}$ & 4,5 & Yes\\
CD --38$\degr$6968 & $<100$ (300) & $<10^{-3}$ & \nodata & \nodata & 
	\nodata & No\\
HD 107146 & 50 -- 250 & $60\pm10$ & $1.5\times10^{-3}$ & \nodata & 
	\nodata & Yes\\
HD 60737 & 50 -- 250 & (300) & $<10^{-3}$ & \nodata & \nodata & No\\
HD 70573 & 50 -- 250 & (300) & $<10^{-3}$ & \nodata & \nodata & No\\
HD 70516 & 100 -- 300 & (300) & $<10^{-3}$ & \nodata & \nodata & No\\
HD 77407 & 20 -- 150 & (300) & $<10^{-3}$ & \nodata & \nodata & No\\
HD 88638 & $250-1000$ & 130 -- 1500 & $1\times10^{-4}-6\times10^{-2}$ 
	& \nodata & \nodata & Yes? \\
HD 17925 & 50 -- 250 & 30 -- 250 & $4\times10^{-6}-3\times10^{-4}$ & 
	$1.3\times10^{-4}$ & 6 & Yes\\
HD 216803 & $200\pm100$ & (300) & $<5\times10^{-4}$ & \nodata & \nodata 
	& No\\
CD --40$\degr$8434 & $\lesssim10$ & $270\pm20, 45\pm5$ & 0.24 & 0.34 & 
	7 & Yes\\
RE J0137+18A & 30 -- 55 & (300) & $<10^{-3}$ & \nodata & \nodata & No\\
RE J0137+18B & 30 -- 55 & (300) & $<10^{-3}$ & \nodata & \nodata & No\\
AP 93 & 80 & (300) & $<7\times10^{-3}$ & \nodata & \nodata & No\\
AA Tau & 1 & $1000\pm100, 190\pm10, 40\pm5$ & 0.51 & 0.31 & 8,9 & Yes\\
DI Tau & 0.6 & (300) & $<3\times10^{-3}$ & \nodata & \nodata & No\\
CD --33$\degr$7795 & $10\pm5$ & (300) & $<10^{-3}$ & \nodata & \nodata 
	& No\\
Hen 3--600A & $10\pm5$ & $260\pm10, 30\pm10$ & 0.16 & 0.21 & 3 & Yes\\
\enddata
\tablenotetext{\dag}{For the stars without detected excesses, the
listed upper limits of $f_d$ are for assumed 300~K debris disks.}
\tablerefs{1.\ \citet{bac97}, M. Meyer et al., in prep.;
2.\ \citet{jur91}; 3.\
\citet{wei02}; 4.\ \citet{jay01}; 5.\ \citet{spa01}; 6.\ \citet{hab01}; 7.\
\citet{gre02}; 8.\ \citet{bec90}; 9.\ \citet{chi01}.}
\end{deluxetable}

\clearpage

\begin{deluxetable}{lcccccccccccc}
\tabletypesize{\scriptsize}
\rotate
\tablewidth{0pt}
\tablecaption{Derived disk properties \label{tab_disks}}
\tablehead{\colhead{Star} & \colhead{$a_{\rm min}$} & 
	\colhead{$\langle a \rangle$} & \multicolumn{3}{c}{$D$ [AU]} &
	\multicolumn{3}{c}{Minimum $M_{\rm dust}$
	[$M_\earth$]\tablenotemark{\dag}} & 
	\multicolumn{3}{c}{$t_{\rm PR}$ [Myr]} & 
	\colhead{Minimum $M_{\rm PB}$} \\ 
	 & [$\micron$] & [$\micron$] & 
	\multicolumn{3}{c}{\hrulefill} & \multicolumn{3}{c}{\hrulefill} &
	\multicolumn{3}{c}{\hrulefill} & [$M_\earth$]\tablenotemark{\dag} \\
	 & & & BB & GB & ISM & BB & GB & ISM & BB & GB & ISM}
\startdata
HD 145519 & 5.7 & 9.4 & 700--30 & 2500--50 & 10$^5$--10$^3$ & 
	0.5--0.001 & 6.5--0.003 & 4000--0.08 & 
	130--0.23 & 1600--0.6 & 10$^6$--10$^2$ & 0.009--0.027 \\
HD 109573A & 3.3 & 5.5 & 37 & 93 & 1100 & 0.14 & 0.83 & 120 & 0.38 & 2.4 & 
	340 & 2.7 \\
HD 102647 & 1.8 & 3.1 & 30 & 100 & 840 & 1.6$\times$10$^{-4}$ & 
	1.7$\times$10$^{-3}$ & 0.12 & 0.32 & 3.5 & 250 & 0.025 \\
HD 107146 & 0.36 & 0.60 & 28 & 290 & 890 & 2.0$\times$10$^{-3}$ & 
	0.22 & 2.0 & 0.48 & 51 & 480 & 0.85 \\
HD 88638 & 0.22 & 0.36 & 4--0.03 & 40--0.08 & 70--0.09 & 
	2$\times$10$^{-6}$--6$\times$10$^{-8}$ & 
	2$\times$10$^{-4}$--4$\times$10$^{-7}$ & 
	5$\times$10$^{-4}$--5$\times$10$^{-7}$ & 10$^{-2}$--10$^{-6}$ & 
	1--10$^{-6}$ & 3--10$^{-5}$ & 0.09--50 \\
HD 17925 & 0.10 & 0.16 & 50--0.8 & 1000--7 & 2000--7 & 
	5$\times$10$^{-6}$--9$\times$10$^{-8}$ & 
	5$\times$10$^{-3}$--7$\times$10$^{-6}$ &
	1$\times$10$^{-2}$--7$\times$10$^{-6}$ & 
	2.0--0.0005 & 2000--0.04 & 4000--0.04 & 
	2$\times$10$^{-4}$--2$\times$10$^{-2}$ \\
CD --40$\degr$8434 & \nodata & \nodata & $\geq$0.85 & $\geq$6.9 & \nodata &
	\nodata & \nodata & \nodata & \nodata & \nodata & \nodata & \nodata \\
AA Tau\tablenotemark{\ddag} & \nodata & \nodata & $\geq$0.077 & $\geq$9.1 &
	\nodata & \multicolumn{3}{c}{1.8$\times$10$^5$} & \nodata & \nodata & 
	\nodata & \nodata \\
Hen 3--600A & \nodata & \nodata & $\geq$0.96 & $\geq$6.3 & \nodata & \nodata &
	\nodata & \nodata & \nodata & \nodata & \nodata & \nodata \\
\enddata
\tablenotetext{\dag}{The assumed mass of the Earth is 
$M_\earth = 5.9736 \times 10^{27}$~gram.}
\tablenotetext{\ddag}{Total dust+gas mass from \citet{bec90,chi01}.}
\end{deluxetable}

\clearpage

\begin{deluxetable}{lcccc}
\tabletypesize{\small}
\tablewidth{0pt}
\tablecaption{Assumed stellar parameters \label{tab_starpars}}
\tablehead{\colhead{Star} & \colhead{Adopted $T_{\rm eff}$} & 
	\colhead{$R_*$} & \colhead{$L_*$} & \colhead{$M_*$} \\ 
	& \colhead{(K)} & \colhead{($R_\sun$)} & \colhead{($L_\sun$)} & 
	\colhead{($M_\sun$)}}
\startdata
HD 145519\tablenotemark{1} & 11000 & 2.2 & 63 & 2.6 \\
HD 109573A & 10000 & 2.0 & 34 & 2.4 \\
HD 102647 & 8500 & 1.8 & 15 & 1.9 \\
HD 107146 & 6000 & 1.2 & 1.7 & 1.1 \\
HD 88638 & 5500 & 1.0 & 0.88 & 0.94 \\
HD 17925 & 5000 & 1.1 & 0.37 & 0.89 \\
CD --40$\degr$8434\tablenotemark{2} & 4400 & 1.39 & 0.64 & \nodata \\
AA Tau\tablenotemark{3} & 4000 & 2.1 & 0.98 & 0.67 \\
Hen 3--600A\tablenotemark{4} & 3200 & 2.8 & 0.72 & \nodata \\
\enddata
\tablenotetext{1}{$T_{\rm eff}$ and $L_*$ from \citet{deg89}.  $T_{\rm
eff}$ is rounded to the nearest 500~K for compatibility with
\citet{kur79} models.}
\tablenotetext{2}{Stellar parameters from \citet{gre02}.}
\tablenotetext{3}{Stellar parameters from \citet{bec90}.}
\tablenotetext{4}{Stellar parameters estimated assuming bolometric correction
of --3.24 \citep[for M4 spectral type;][]{flo96}, and 
$V$=12.04 \citep{tor00}.}
\end{deluxetable}

\end{document}